\documentclass[AMA,STIX1COL]{WileyNJD-v2}

\usepackage{graphicx}
\usepackage{bm}
\usepackage{graphicx}
\usepackage{subfigure}
\usepackage{amsmath}
\usepackage[T1]{fontenc}
\usepackage{color}
\usepackage{url}

\articletype{Article Type}%

\received{xxx}
\revised{xxx}
\accepted{xxx}

\begin{document}

\title{Effects of particle size-shape correlations on shear strength of granular materials: The case of particle elongation}

\author[1,2]{Sergio Carrasco}

\author[1,2]{David Cantor}

\author[1,2]{Carlos Ovalle*}

\authormark{Sergio Carrasco \textsc{et al}}

\address[1]{\orgdiv{Department of Civil, Geological and Mining Engineering}, \orgname{Polytechnique Montr\'eal}, \orgaddress{\state{Qu\'ebec}, \country{Canada}}}

\address[2]{\orgdiv{Research Institute of Mining and Environment (RIME)}, \orgname{UQAT-Polytechnique}, \orgaddress{\state{Qu\'ebec}, \country{Canada}}}

\corres{*Corresponding author name. \email{carlos.ovalle@polymtl.ca}}


\abstract[Summary]{
  Granular materials often present correlations between particle size and shape due to their geological formation and mechanisms of weathering and fragmentation. 
  It is known that particle shape strongly affects shear strength. 
  However, the effects of shape can be modified by the role the particle plays in a sample given its size. 
  We explore the steady shear strength of samples composed of particles presenting size-shape correlations and we focus on the case of particle elongation in two opposite scenarios: (A) large elongated grains with finer circular grains and (B) large circular grains with elongated finer grains. 
  By means of numerical simulations, we probe the shear strength of samples of varying particle size span from mono to highly polydisperse and particle aspect ratios varying between 1 and 5. 
  We find that the two correlations tested strongly impact the shear strength as particle size span evolves. 
  Microstructural analyses allow us to identify how each correlation affects  connectivity and anisotropies linked to the orientation of the particles and load transmission. 
  Decompositions of the stress tensor let us identify the sources of the different mechanical behavior in each correlation and determine the contributions of each particle shape to macroscopic shear strength. 
  This study proves that common small-scaling methods based on truncated or parallel particle size distributions can incur in under/over-estimations of shear strength if particle shapes are not considered in the scaling process. }



\maketitle

\section{Introduction}

The mining industry generates large amounts of loose uncompacted granular materials, such as ore stockpiles and mine waste rock dumps that can reach several hundred meter high \cite{Linero2007,Valenzuela2008,Bard2012,Aubertin2013}. 
Physical stability of these large structures must be verified all through engineering designs, because any mass sliding could significantly affect the environment and the mining operation \cite{Santamarina2019,Aubertin2021}. 
Slope stability of loose granular fills depends on the geometry of the structure, the loading conditions, and the mechanical properties of the granular media \cite{Duncan2014}. 
For the latter, mechanical laboratory tests are typically carried out to fit steady-state strength (SSS) parameters for a given failure criterion. 

It is well known that normalized SSS (or critical friction angle) of loose granular materials does not depend on packing density, but only on the properties of the grains \cite{Biarez1994}. 
Recent works have shown that if the grains have the same shape, SSS does not depend on the particle size distribution (see numerical results in Refs. \cite{Voivret2007,Voivret2009,MuirWood2008,Azema2017,Cantor2018,Linero2019}, and experimental results in Refs. \cite{Li2013,Yang2018}). 
This finding implies that, since particle shape and grain roughness remain unchanged along with different grain sizes for a given material, representative SSS can be obtained through shearing tests of samples with different particle size distribution (psd). 
For instance, SSS of very coarse granular materials, such as rockfills or mine waste rock, can be measured through shearing tests of small-scaled samples after removing the coarse (oversized) fraction that cannot be handled in laboratory devices. 
Among small-scaling methods we can find, for instance, scalping, replacement, or parallel gradation \cite{Marachi1969,Verdugo2007,Ovalle2014,Ovalle2020,Deiminiat2020}. 

On the other hand, several studies have demonstrated that the SSS strongly depends on particle shape, and materials composed of more angular and/or elongated grains have higher strengths than rounded ones \cite{Cho2006, Azema2010,Matsushima2011,Nguyen2015,Altuhafi2016,Xiao2019,Linero2019,Sarkar2019,Xu2021}. 
Assemblies of elongated angular particles have high heterogeneity of void size distribution, which enhances grain interlocking and particles tend to slide instead of roll, resulting in higher strength \cite{Yang2012,Pena2007}. 

Particle shape effects could then potentially invalidate small-scaling methods based on altering the psd in granular materials, for instance, composed of rock clasts with strong particle size-shape correlation; i.e., grain shape varies with grain size. 
For instance, \cite{Linero2017} reported a mine waste rock composed of colluvial grains weathered from anisotropic sedimentary rocks, where the coarser the rock aggregate, the more flat the grain shape.
This phenomenon was linked to preferential weaknesses in the material due to lamination on large particles. 
On the contrary, \cite{Ovalle2020a} characterized a rockfill material from shale rock where metamorphism induced fine foliation producing more elongated particles as their size was small. 
In those materials, altering the psd will necessarily change the characteristic shape of the particles, thus affecting the SSS.
Therefore, to capture the SSS of prototype coarse materials using small-scaled samples, one should consider not only the size of the grains, but also a representative distribution of particle shapes. 
However, while particle shape has been largely studied, the effect of shape varying with different particle sizes in a given material remains unclear. 

The main objective of this paper is to study the combined effects of particle size and particle shape distributions on the SSS of granular media. 
In particular, we are interested in the correlation between size and elongation of the grains. 
By means of two-dimensional discrete-element simulations, we systematically study samples with increasing particle size span linked to the elongation of the grains. 

This paper is organized as follows. 
In Sec. \ref{sec:numerical}, we introduce the particle size-shape correlation and the numerical approach to build and shear 2D samples over a large range of grain sizes and grain shape elongations. 
In Sec. \ref{sec:macro}, we analyze the macromechanical behavior and packing properties in terms of solid fraction and shear strength as a function of shear deformation up to the steady state. 
In Sec. \ref{sec:micro}, we use microstructural descriptors to characterize the effect of size-shape correlations on particle orientation and connectivity at the SSS. 
Section \ref{sec:roles} is focused on the analysis of contributions to strength of micromechanical parameters linked to the geometrical configuration of particles and force transmission mechanisms. 
We also identify the role of each shape class has on the macroscopic strength.
Finally, we conclude and draw perspectives of this work. 

\section{Model material and numerical simulation}\label{sec:numerical}
\begin{figure}[h]
  \centering
  \includegraphics[width=0.45\linewidth]{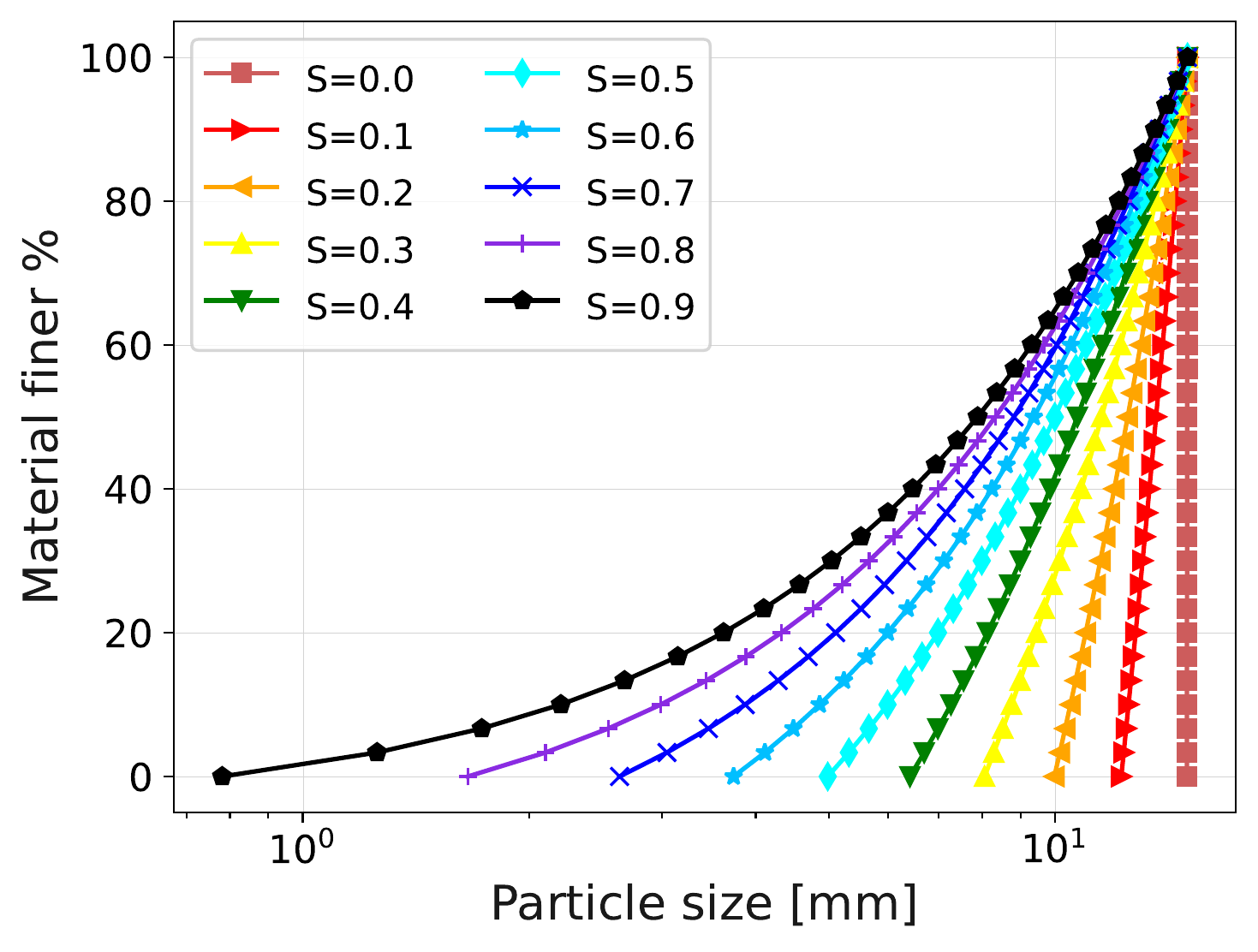}
  \caption{Particle size distributions (psd) as a function of particle size span $S$. The distributions are uniform by volume fractions.}
  \label{fig_1_2}
\end{figure}

The correlation shape-size of granular materials is modeled in this study by using two-dimensional particles of circular or elongated shapes. 
First, we set the dispersity of particle sizes using parameter $S$ defined as 
\begin{equation}\label{eq:S}
S = \frac{d_{max} - d_{min}}{d_{max} + d_{min}},
\end{equation}
with $d_{max}$ and $d_{min}$ being the maximal and minimal particle diameter, respectively, in a sample. 
We varied $S$ in range $0$ to $0.9$ in steps of $0.1$, which means that we considered monodisperse samples, up to polydisperse configurations in which $d_{max}/d_{min} = 19$. 
For simplicity, we define a uniform particle size distribution (psd) by volume fractions between $d_{min}$ and $d_{max}$. 
The resulting psd are presented in Fig. \ref{fig_1_2} for the different values of particle size span $S$. 

\begin{figure}[h]
  \centering
  \subfigure[ ] {
  \includegraphics[width=0.25\linewidth]{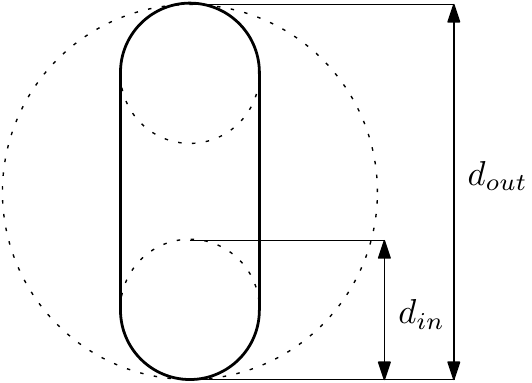}}
  \subfigure[ ] {
  \includegraphics[width=0.25\linewidth]{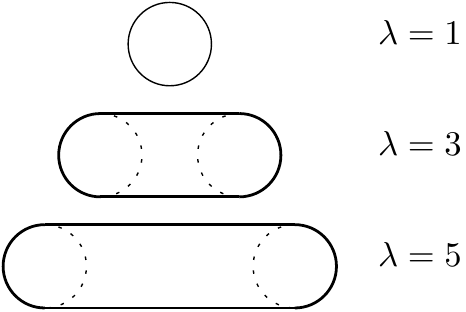}}
  \caption{(a) Scheme presenting the inner and outer circles of an elongated particle, and (b) examples of particles with elongation equal to $1$, $3$, and $5$.}
  \label{fig:schemeparticles}
\end{figure}

For the elongated particles, we consider rounded-cap rectangles, as shown in Fig. \ref{fig:schemeparticles}(a).
The size of these particles, associated with the psd, is the shortest dimension $d_{in}$ (i.e., the inscribed circle within the particle).
The elongation of these particles is then defined using parameter $\lambda$ as 
\begin{equation}
\lambda = d_{out}/d_{in},
\end{equation}
where $d_{out}$ is the circumscribed circle around the particle. 

Since the genesis of granular materials can either produce rounded or elongated particles as they weather or break, as mentioned above, we need to explore two different scenarios. 
For the case in which primary particles are elongated and smaller grains tend to be more rounded, we set the coupled evolution size-shape to follow the next equation
\begin{equation}\label{eq:caseA}
\lambda(d) = \lambda_{min} + \frac{d-d_{min}}{d_{max} - d_{min}} (\lambda_{max} - \lambda_{min}),
\end{equation}
and, for the case in which primary particles are rounded and smaller ones are elongated, we have the complementary relation
\begin{equation}\label{eq:caseB}
\lambda(d) = \lambda_{min} + \frac{d_{max}-d}{d_{max} - d_{min}} (\lambda_{max} - \lambda_{min}),
\end{equation}
in which the aspect ratio $\lambda$ of a particle is fully defined by the size $d$ for a given value of particle size span $S$. 
For simplicity, we name these two correlations in Eq. (\ref{eq:caseA}) and (\ref{eq:caseB}), \emph{case A} and \emph{case B}, respectively.
We also fixed the extreme aspect ratio values for all samples with $\lambda_{min} = 1$ and $\lambda_{max} = 5$ (see some examples in Fig. \ref{fig:schemeparticles}(b)). 
This means that each sample presents all ranges of $\lambda$ between the size limits mentioned before, and not only broader particle size distributions are able to present varied particle shapes. 
The correlations produce samples where the volume fractions across particle sizes are kept constant for different particle shapes. 
In other words, our samples can be considered as a scaling approach in which particle shapes are transferred from wide particle size distributions to narrower size limits. 

\begin{figure}[htb]
    \centering
    \subfigure[ ] {
    \includegraphics[width=0.23\linewidth]{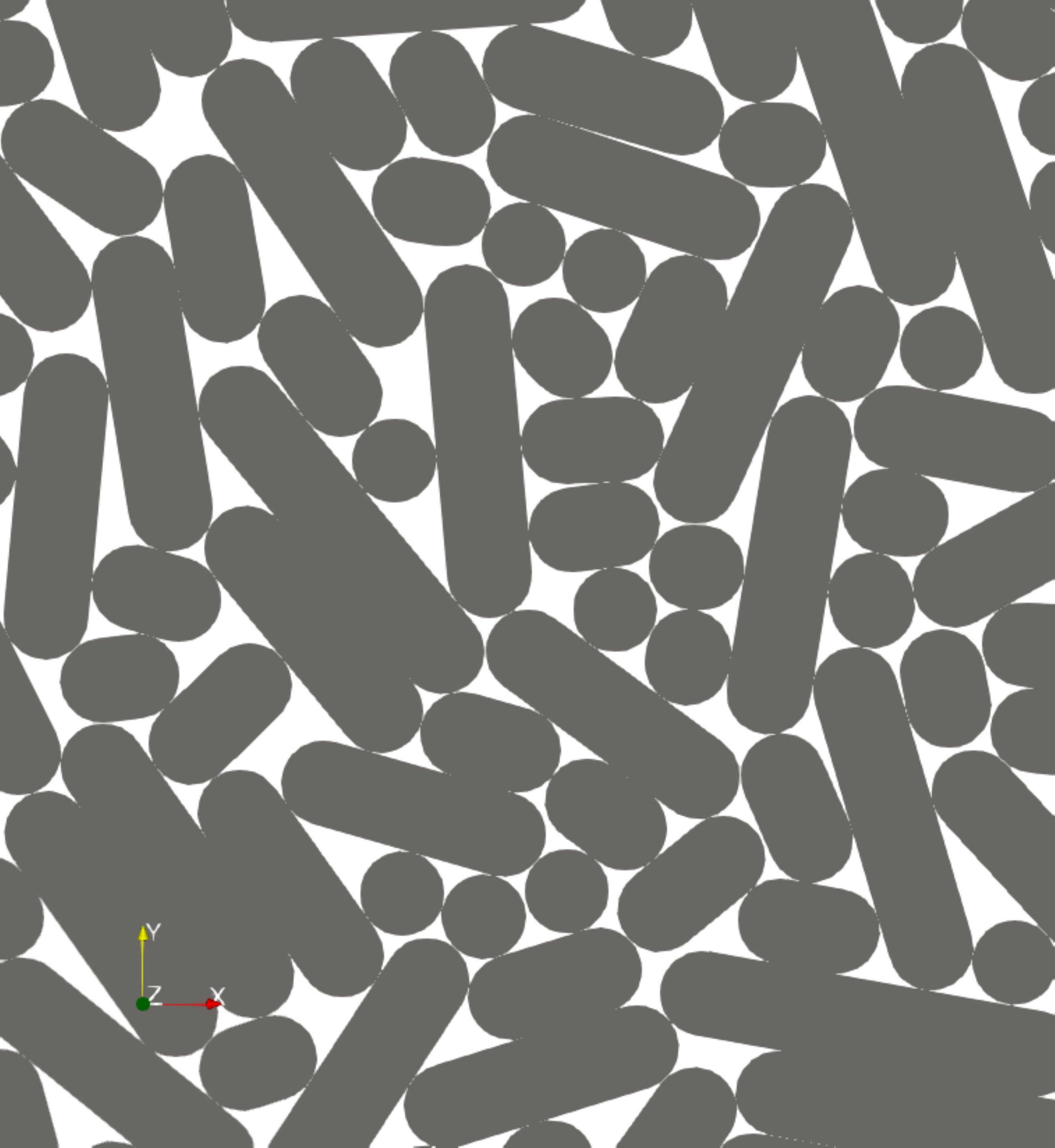}}
    \subfigure[ ]  {
    \includegraphics[width=0.23\linewidth]{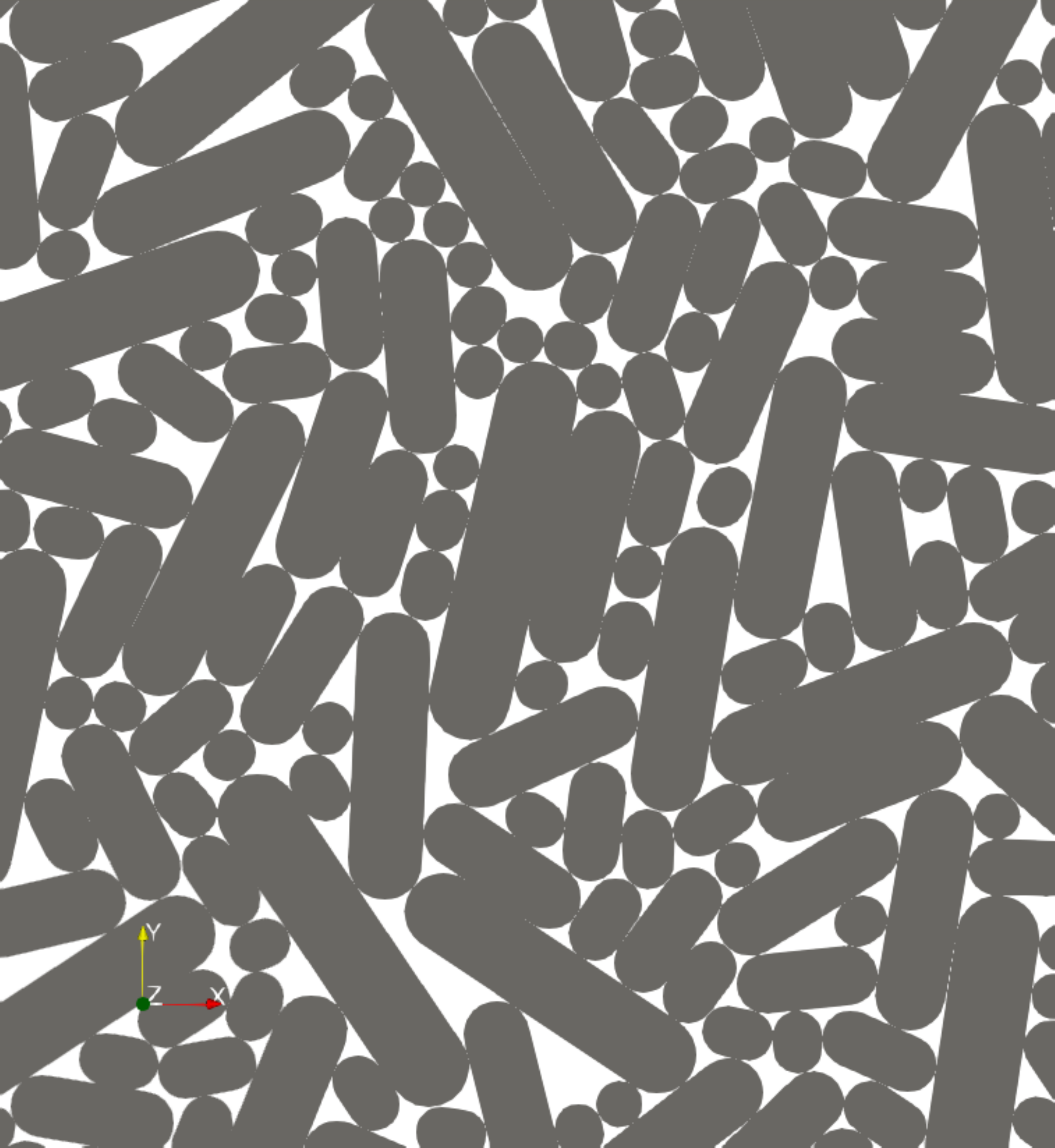}}
    \subfigure[ ]  {
    \includegraphics[width=0.23\linewidth]{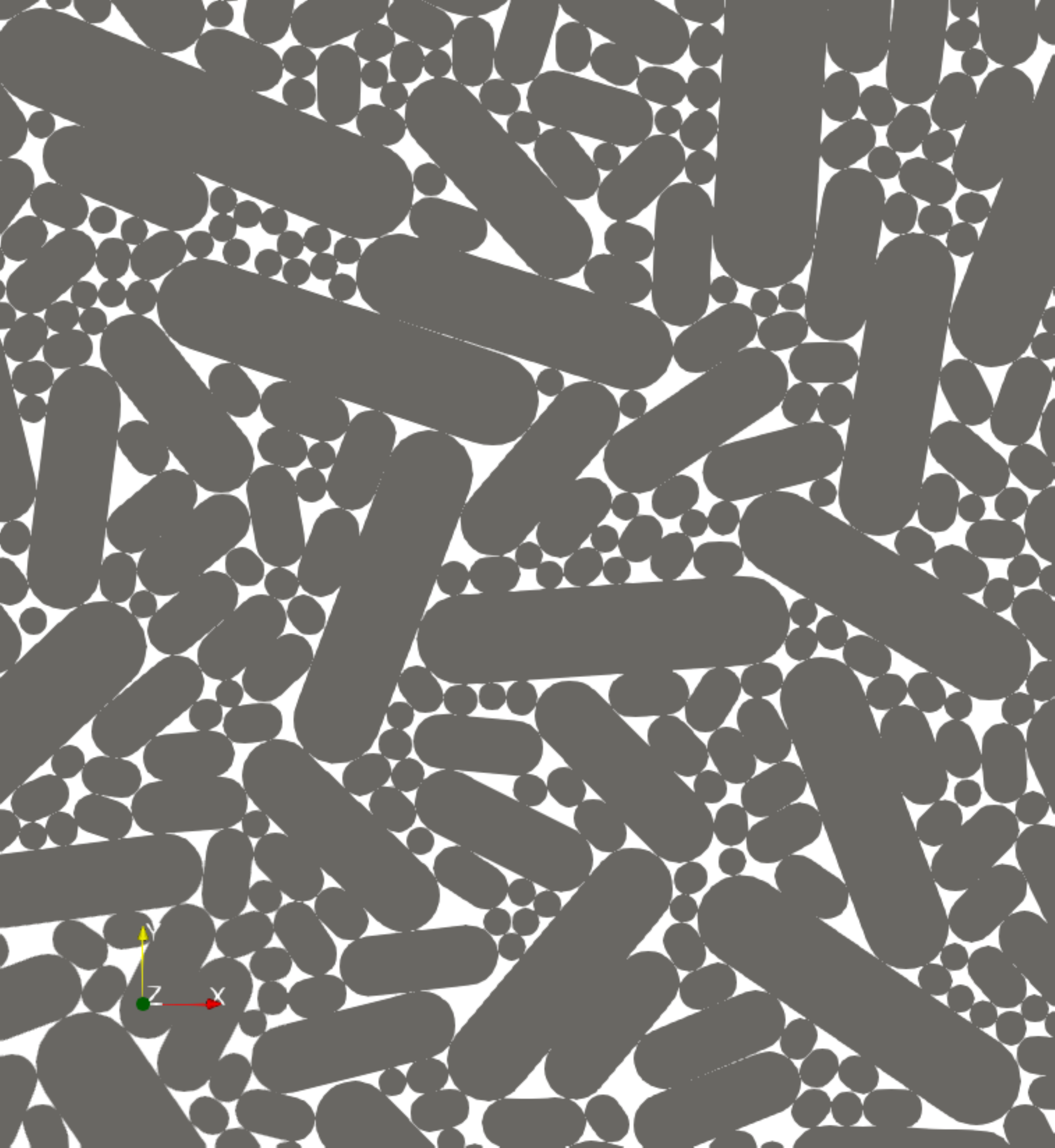}}
    \subfigure[ ]  {
    \includegraphics[width=0.23\linewidth]{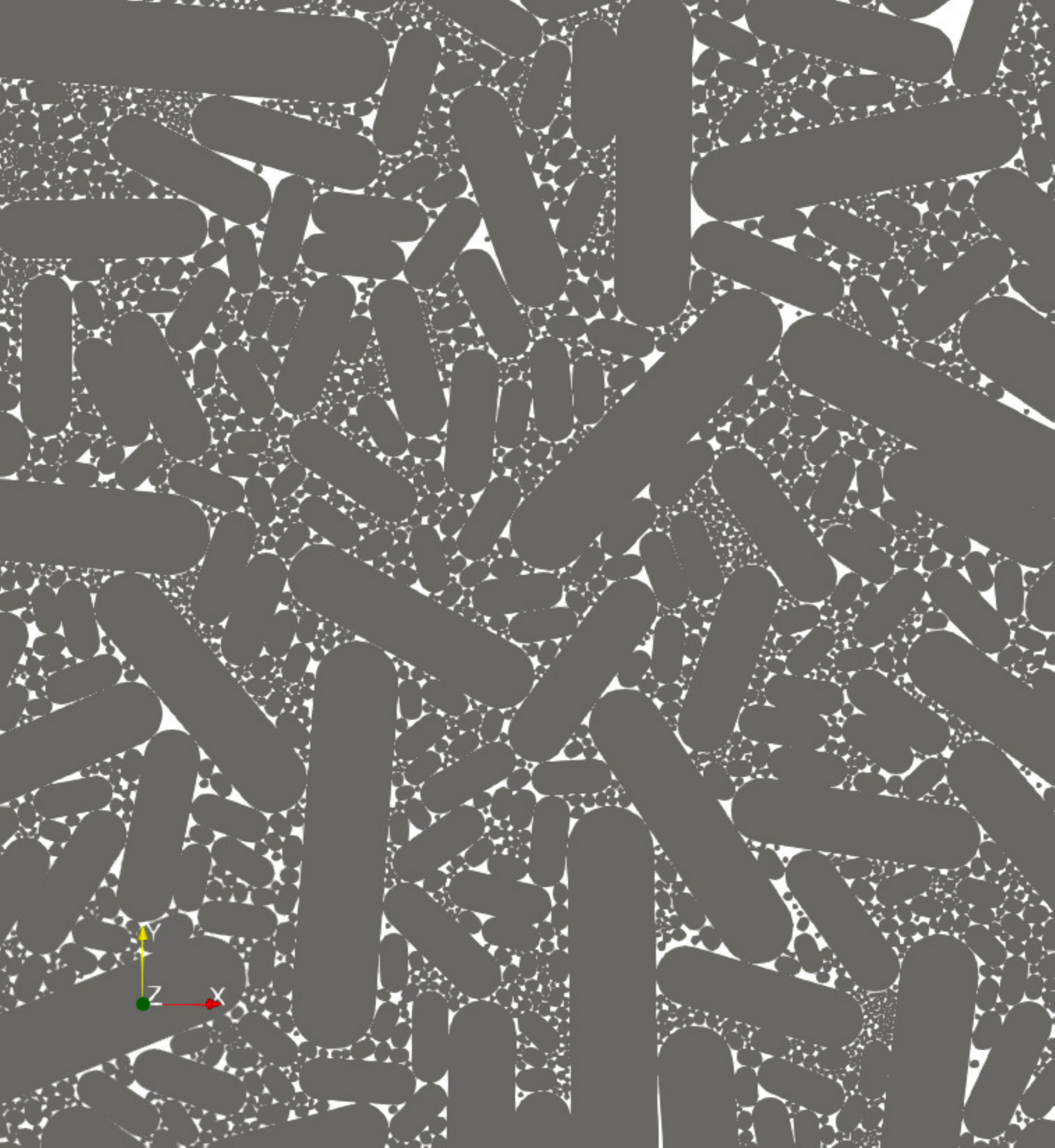}}
    \subfigure[ ] {
    \includegraphics[width=0.23\linewidth]{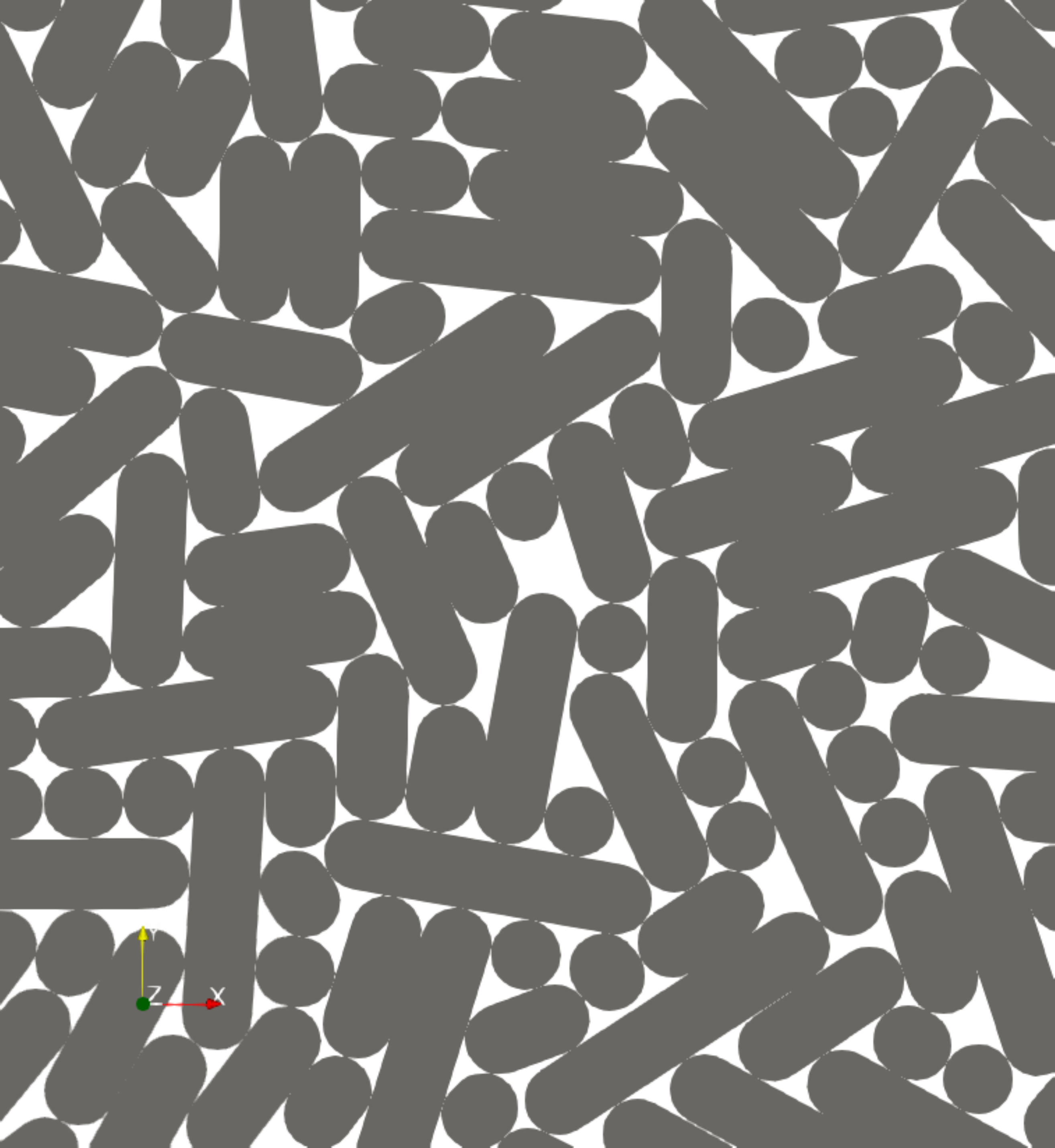}}
    \subfigure[ ]  {
    \includegraphics[width=0.23\linewidth]{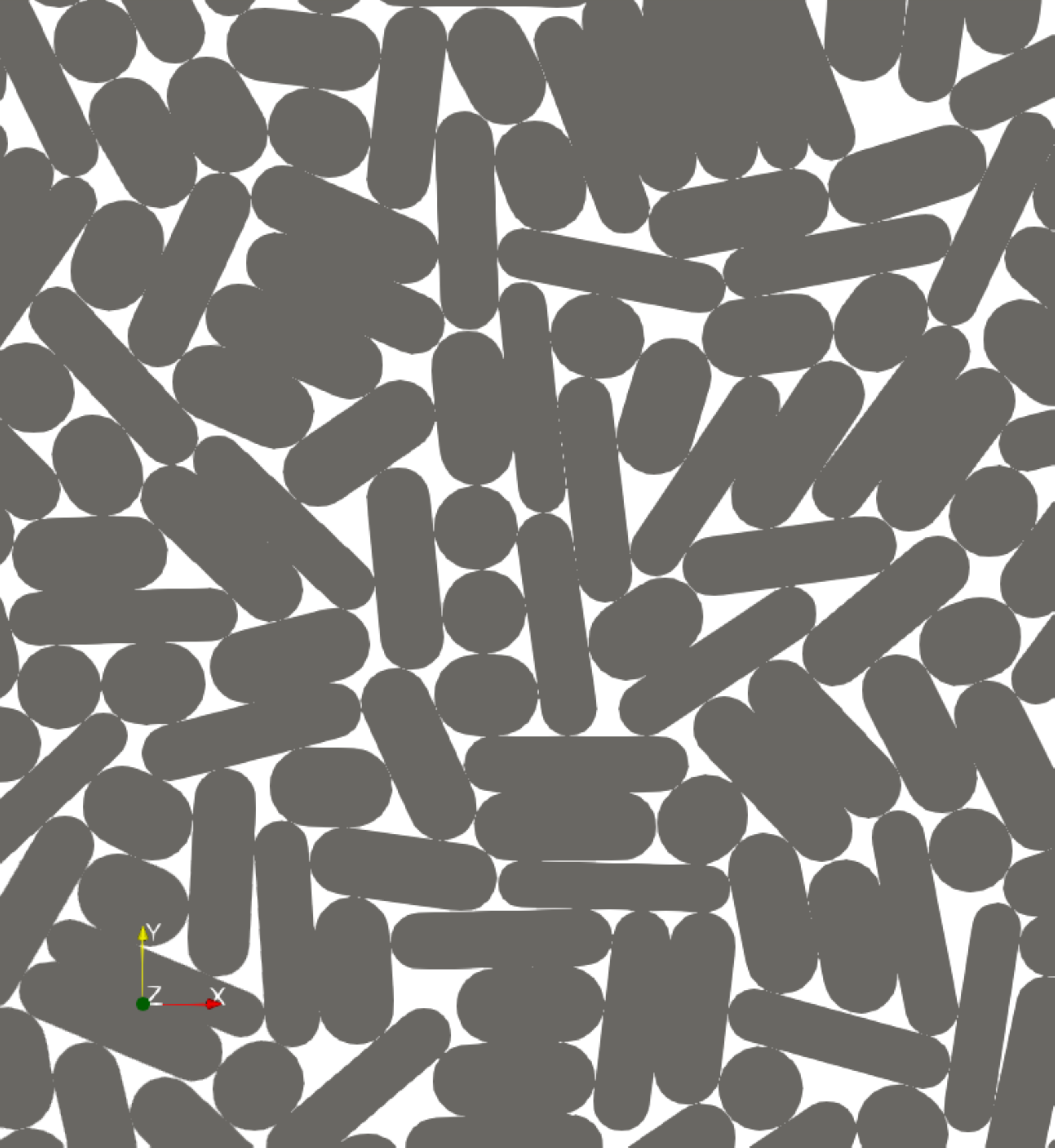}}
    \subfigure[ ]  {
    \includegraphics[width=0.23\linewidth]{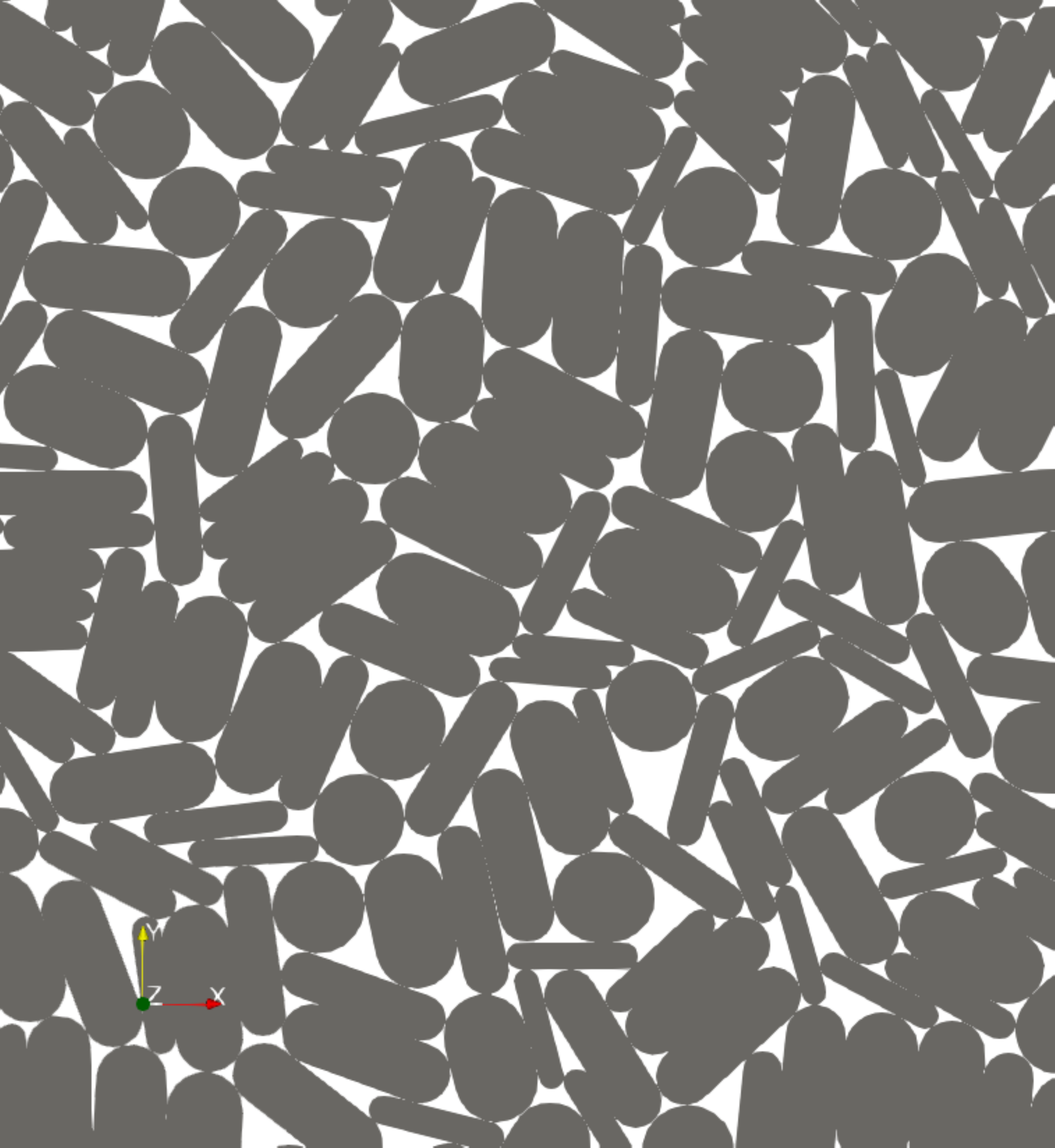}}
    \subfigure[ ]  {
    \includegraphics[width=0.23\linewidth]{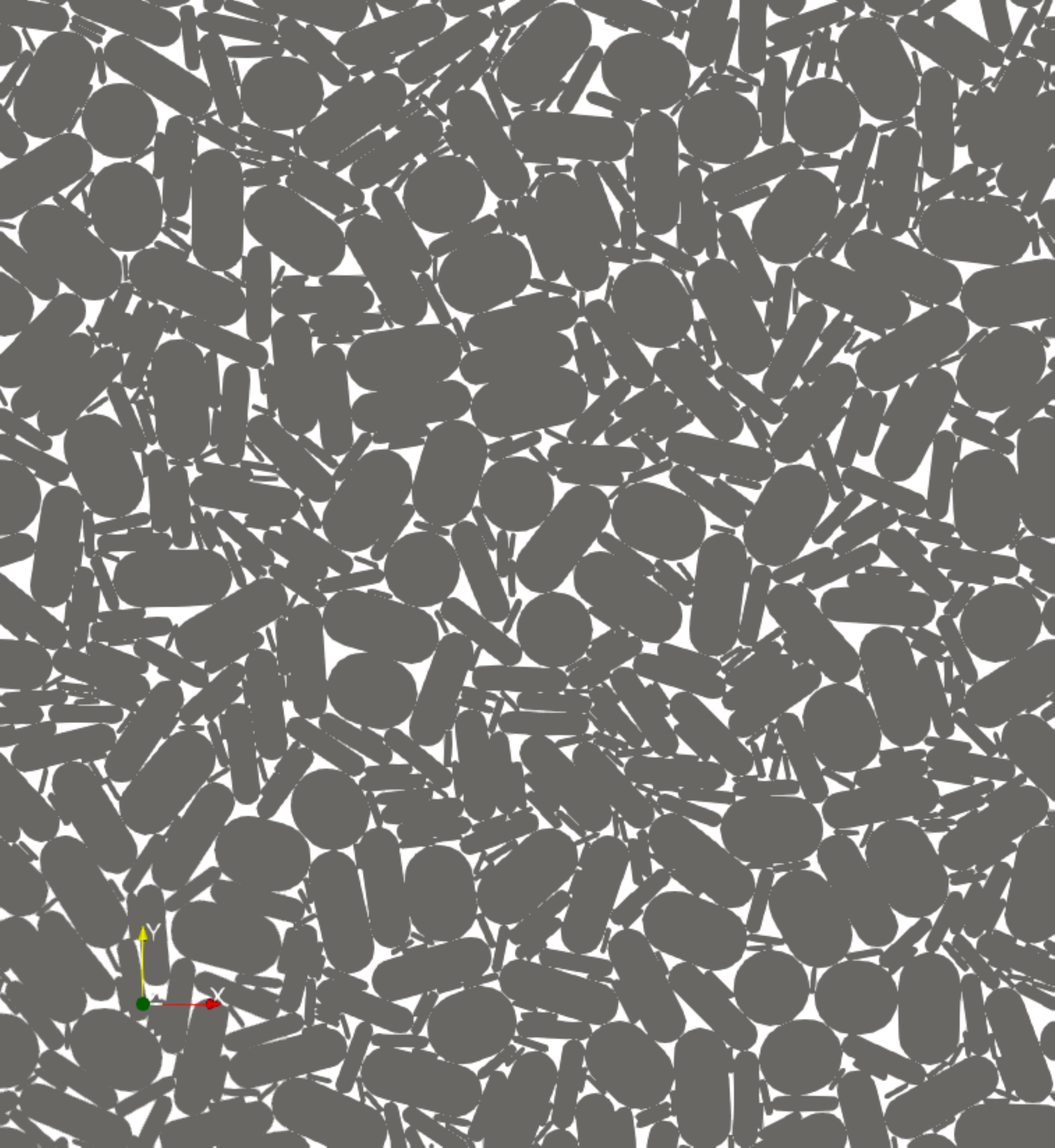}}
  \caption{Screenshots of samples for case A (top) and case B (bottom) for particle size spans $S = 0.0$ (a, b), $S = 0.3$ (c, d), $S = 0.6$ (e, f), and $S = 0.9$ (g, h).}
    \label{fig_cases}
\end{figure}

Our samples are made up of $N_p \simeq 10 \ 000$ particles that are placed in boxes by using an algorithm of deposition based on geometrical rules. 
We initially compress the samples using an isotropic pressure $P$ via rigid walls around the assemblies up to stable configurations in which the solid fraction $\nu=V_s/V$, being $V_s$ the volume of the grains and $V$ the volume of the sample, presents only fluctuations under the $0.1\%$ of the mean value. 
Figure \ref{fig_cases} presents a set of samples for varying $S$ for cases A and B at the end of this compression step.

\begin{figure}[htb]
  \centering
  \includegraphics[width=0.4\linewidth]{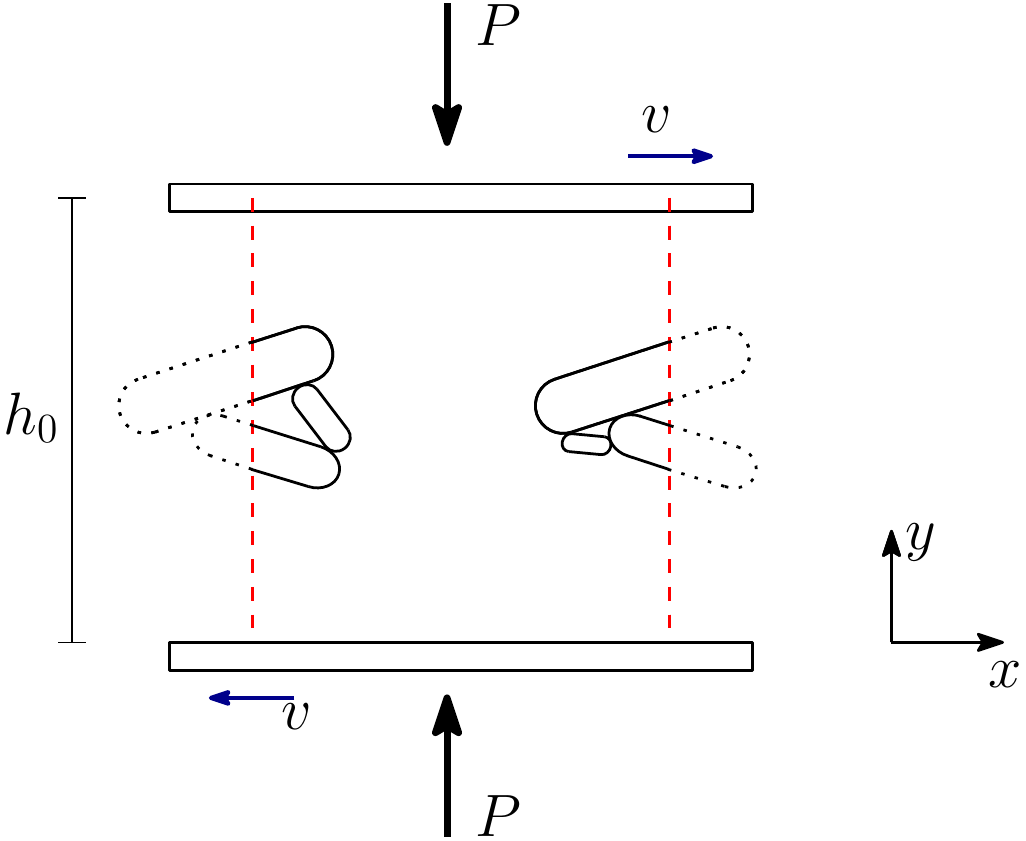}
  \caption{Scheme of boundary conditions for the shearing tests. The dashed lines represent the periodic boundary along the `$x$' axis.}
  \label{fig:boundaries}
\end{figure}

For the shearing tests, we used periodic boundary conditions along the horizontal direction, implying that any particle or contact crossing the boundary will instantaneously reappear on the corresponding opposite side of the sample (see a scheme of boundary conditions in Fig. \ref{fig:boundaries}). 
Then, the shearing tests are undertaken by moving both walls along axis `$x$' at a constant velocity $v$ while applying load $P$ along the axis `$y$'.
The shear velocity $v$ is set to follow a quasi-static flow condition, imposed by the inertial number $I=\dot{\gamma} \langle d \rangle \sqrt{\rho/P} \ll 1$, where $\langle d \rangle$ is the average particle diameter, $\rho$ the particle density, and $\dot{\gamma} = v/h_0$, with $h_0$ being the height of the sample at the beginning of the test. 
In all our simulations, the inertial number was set to $I=1\times10^{-3}$. 

Samples composed of elongated particles often require large deformations to reach a steady state, not only in terms of macroscopic descriptors, but also in terms of particle organization (e.g., particle orientation). 
For this reason, we sheared our samples up to a cumulated shear strain $\gamma = 400 \% = \delta/h_0$, with $\delta$ the cumulated displacement of the walls. 

Our simulations were performed using the discrete-element approach known as contact dynamics (CD), which considers collections of rigid bodies interacting with unilateral frictional contacts. 
This strategy employs non-smooth contact laws to iteratively solve contact forces and particle velocities via an implicit time-stepping scheme. 
For more information on the mathematical framework of the CD method or its implementation, please see Refs. \cite{Moreau1988, Jean1992,Radjai2009, Dubois2018}.
We used the platform LMGC90, which is a free, open-source software for the simulation of discrete mechanical systems in the frame of the CD method \cite{LMGC90Web,Dubois2011}. 
In all the tests, the coefficient of friction between particles was set to 0.4 and gravity was neglected. 
During shearing, particles in contact with the walls were preset to follow the wall displacements by coupling their degrees of freedom, so no slip occurs at the interface wall-particles. 
Videos of the shearing tests can be found in the following link \url{https://youtu.be/xI20RGulp6k}. 

\section{Macroscopic behavior}\label{sec:macro}
The macroscopic shearing behavior is characterized using the evolution of the solid fraction $\nu$ and the shear strength $q/p$, being $q$ the deviatoric stress and $p$ the mean pressure of the granular stress tensor $\sigma$. 
This tensor is found using the expression
\begin{equation}\label{eq_stresstensor}
\sigma_{ij} = \frac{1}{V} \sum_{\forall c} f_i^c \ell_j^c
\end{equation}
with $f$ the force and $\ell$ the branch vectors (i.e., the vector joining the center of mass of touching particles at contacts $c$). 
We compute $q=(\sigma_1 - \sigma_2)/2$ and $p=(\sigma_1 + \sigma_2)/2$, with $\sigma_1$ and $\sigma_2$ the principal stresses of $\sigma$.
Note that $q/p = \sin(\phi)$ at the steady state, with $\phi$ being the macroscopic friction angle of the material. 
We characterize the steady state by averaging the different parameters for the last $20\%$ of deformation out of the total $400\%$. 
We observed steady values for both $\nu$ and $q/p$ as early as $\gamma = 250\%$; however, we applied more deformation to let other structural parameters stabilize, in particular, the orientation of elongated particles. 

\begin{figure}[tbh]
  \centering
  \includegraphics[width=0.5\linewidth]{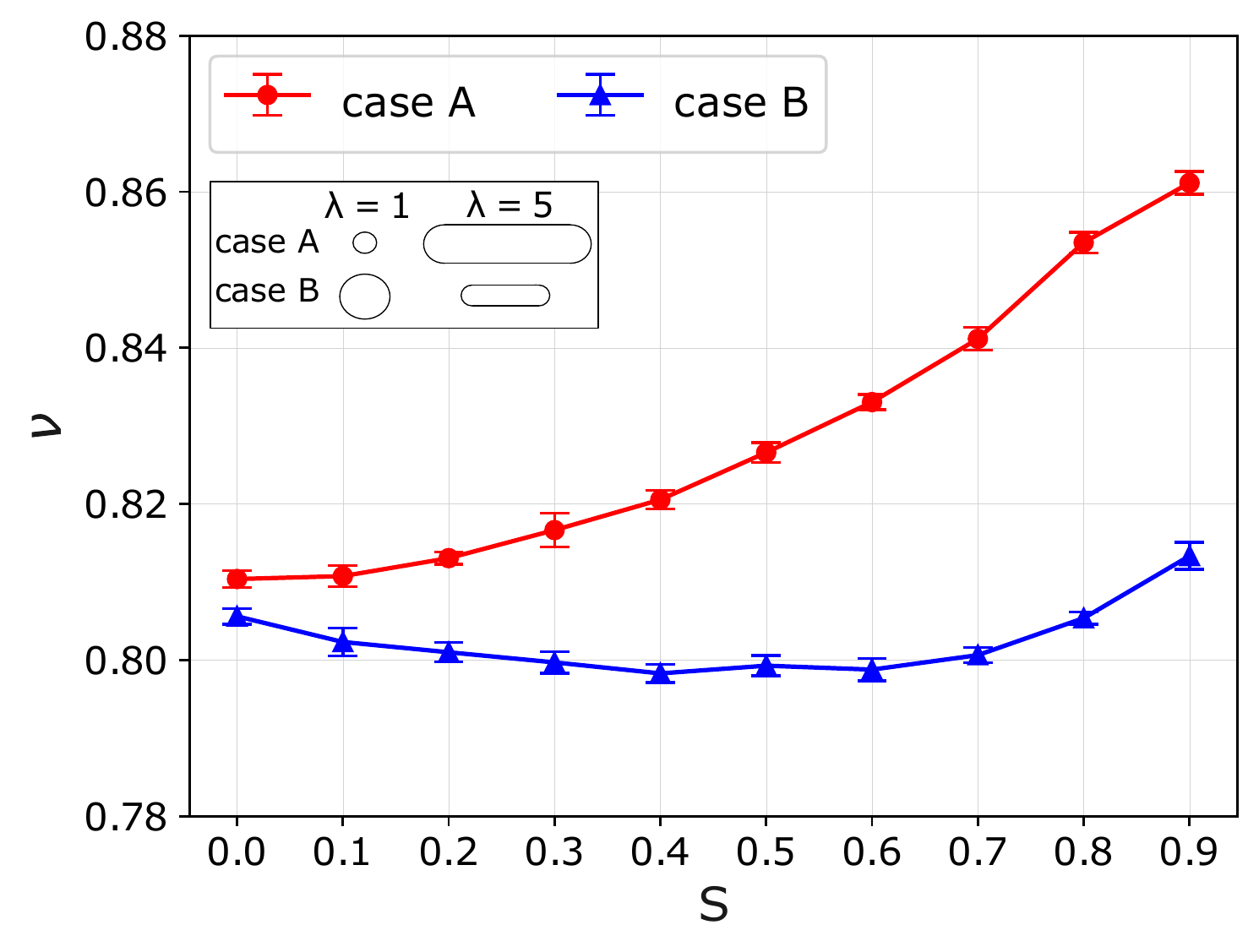}
  \caption{Evolution of the solid fraction $\nu$ at the steady state as a function of the particle size dispersion $S$ and for cases A and B. 
  Error bars display the standard deviation of the data for the last $20\%$ of deformation.}
  \label{fig:ss_solidfraction}
\end{figure}

Figure \ref{fig:ss_solidfraction} shows the evolution of the solid fraction as a function of the particle size dispersity $S$ for cases A and B. 
For case A, we observe that $\nu$ gradually increases with $S$ as the smaller rounded grains are capable of filling the pores left by cavities created by larger elongated grains. 
Surprisingly, case B does not show an analogous behavior. 
In this case, in which the smaller particles present elongated shapes, the solid fraction seems to be only slightly affected by the grain size distribution and varies with a parabolic trend with a minimal solid fraction of $\nu \simeq 0.8$ for $S =0.4$
In other words, these samples are unable to develop large enough cavities to allow the smaller elongated particles to fit in. 
It is only after $S>0.6$ that the size dispersion can create denser configurations. 

\begin{figure}[tbh]
  \centering
  \includegraphics[width=0.5\linewidth]{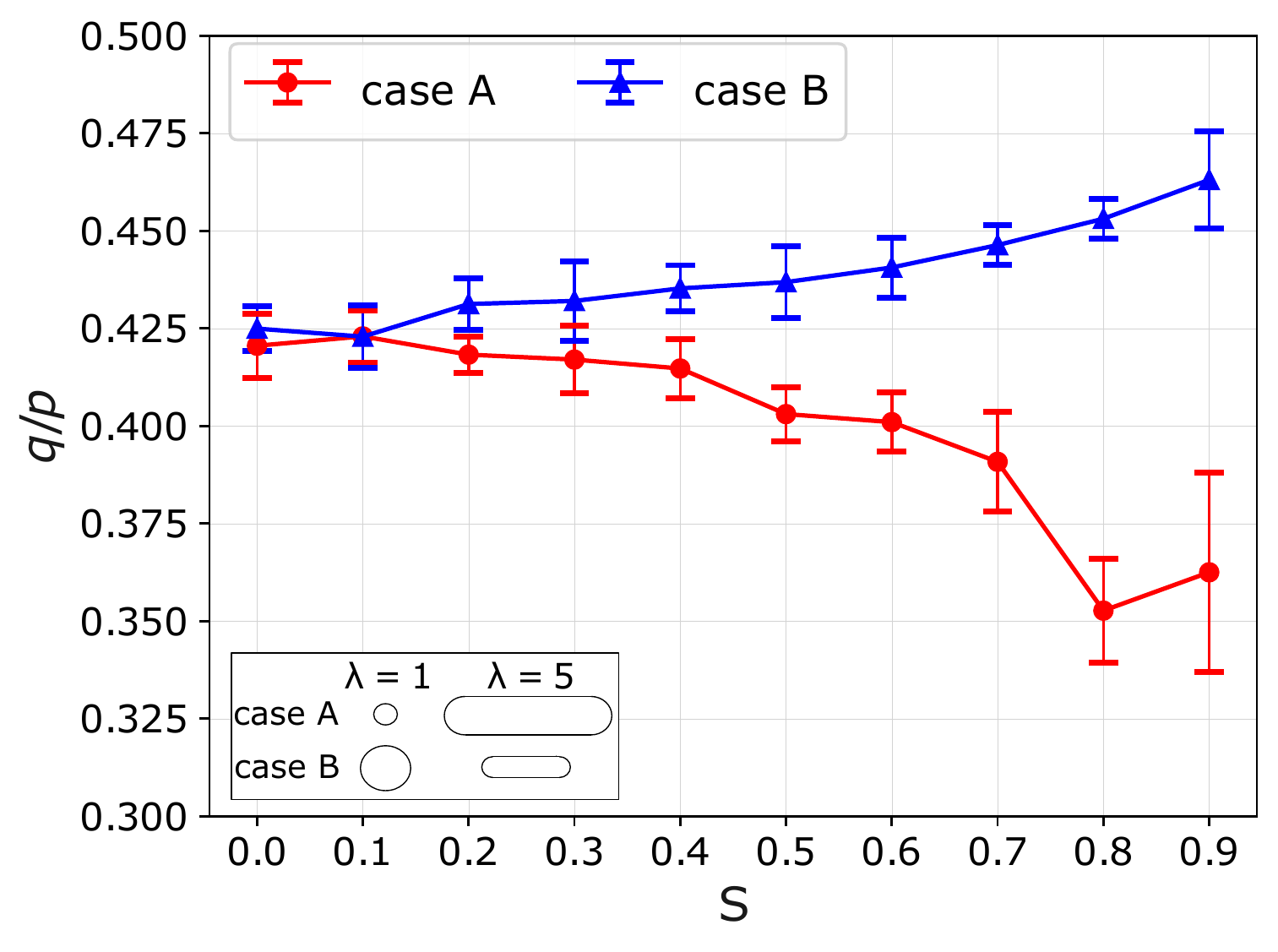}
  \caption{Evolution of the shear strength $q/p$ at the steady state as a function of $S$ for cases A and B.
  Error bars display the standard deviation of the data for the last $20\%$ of deformation.}
  \label{fig:ss_strength}
\end{figure}

In terms of shear strength, Fig. \ref{fig:ss_strength} presents the evolution of $q/p$ with $S$ for cases A and B. 
For case A, we observe that the shear strength gradually decreases with $S$ despite the increase in solid fraction we pointed out in the previous figure. 
For case B, the shear strength barely increases from $q/p \simeq 0.424$ for $S=0$ to $q/p \simeq 0.46$ for $S=0.9$. 
Remarkably, the evolution of density and strength seem uncorrelated for both cases A and B. 

The evolution of $\nu$ and $q/p$ for our set of samples is counterintuitive and diverges from several previous studies showing that particle size distribution does not affect the strength of granular materials. 
In particular, numerical approaches have shown that elongated particles promote an increase of shear strength \cite{Azema2010}, and large particles use to bear a larger proportion of the external load than the small particles \cite{Nguyen2015}. 
That combined size-shape effect does not seem to be reproduced in case A, in which the larger particles are elongated but shear strength turns out to be lower. 
In turn, for case B, the grain size span effect seems practically negligible when considering the density or the shear strength of the samples. 

These results show that particle size-shape correlations deeply modify the steady state strength and density of granular materials. 
Foremost, this also suggests that common scaling methods for coarse granular materials should avoid discarding certain granulometric classes only based on particle size, but should also focus on particle shape representativeness on small-scaled samples.

In order to understand the behavior found in our experiments, we need to explore the microstructural characteristics of the samples and the contributions of each size/shape class to the macroscopic strength. 

\section{Microscopic descriptors}\label{sec:micro}
\begin{figure}[htb]
  \centering
  \subfigure[ ] {\includegraphics[width=0.23\linewidth]{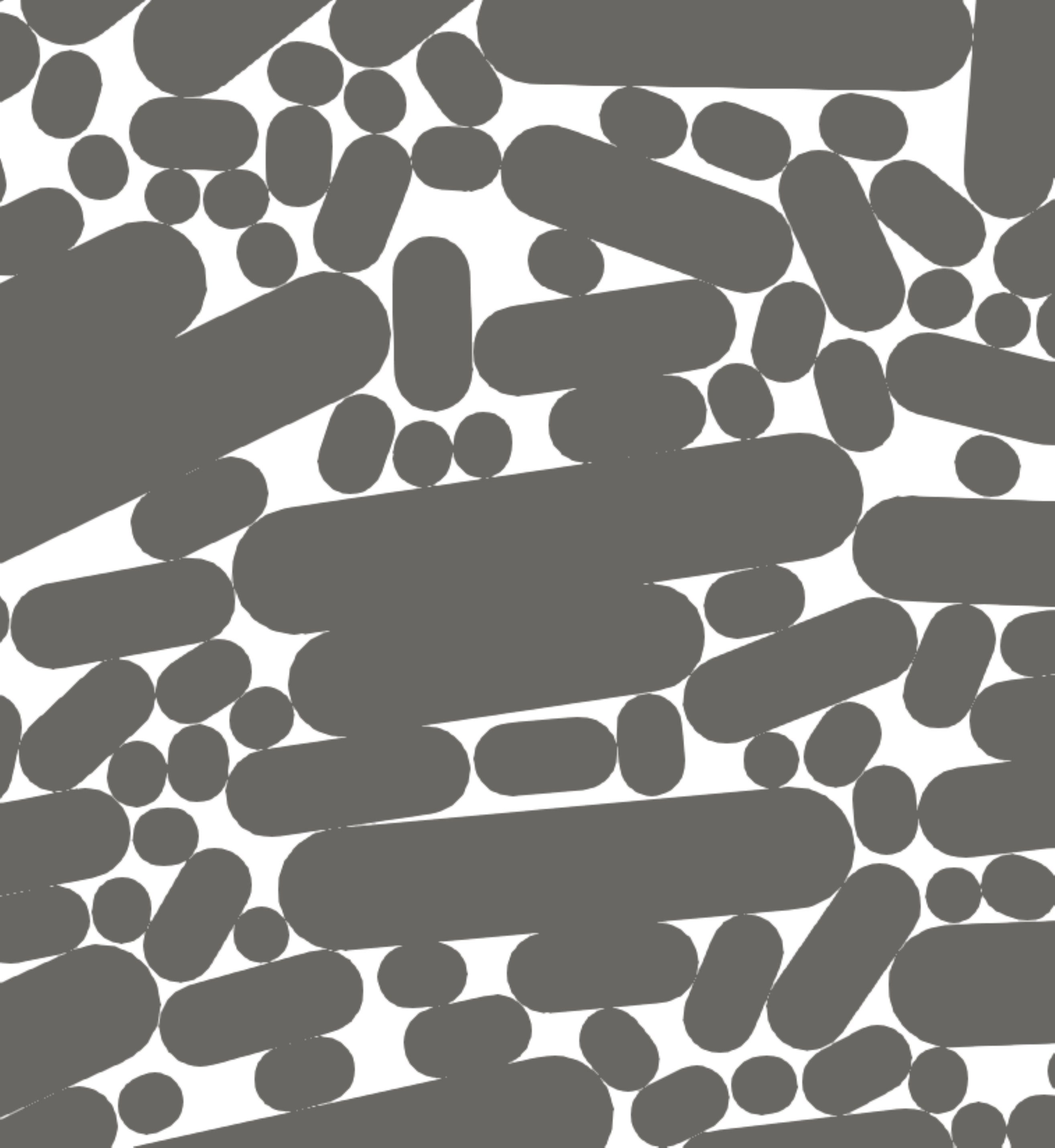}}
  \subfigure[ ] {\includegraphics[width=0.23\linewidth]{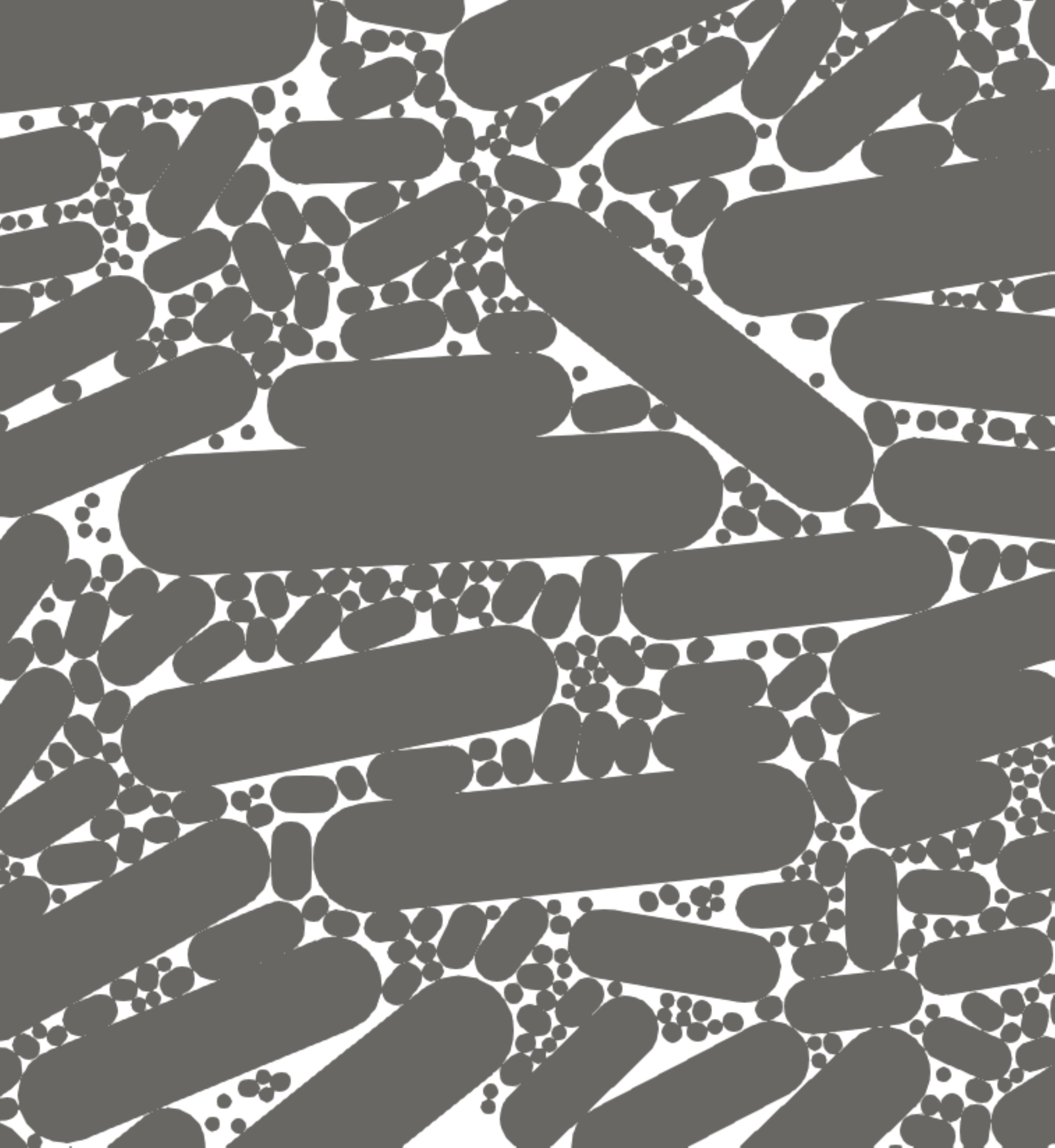}}
  \subfigure[ ] {\includegraphics[width=0.23\linewidth]{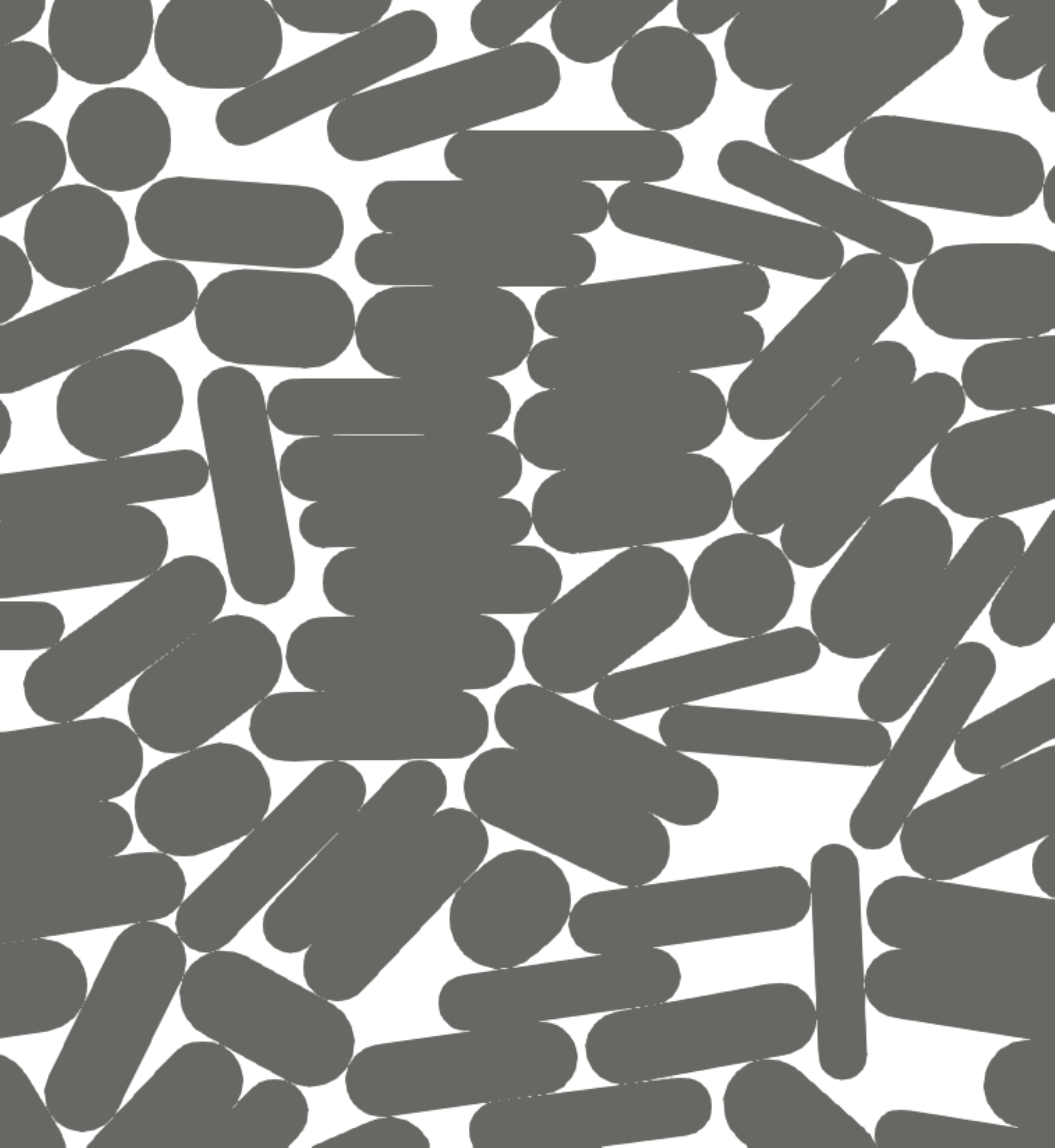}}
  \subfigure[ ] {\includegraphics[width=0.23\linewidth]{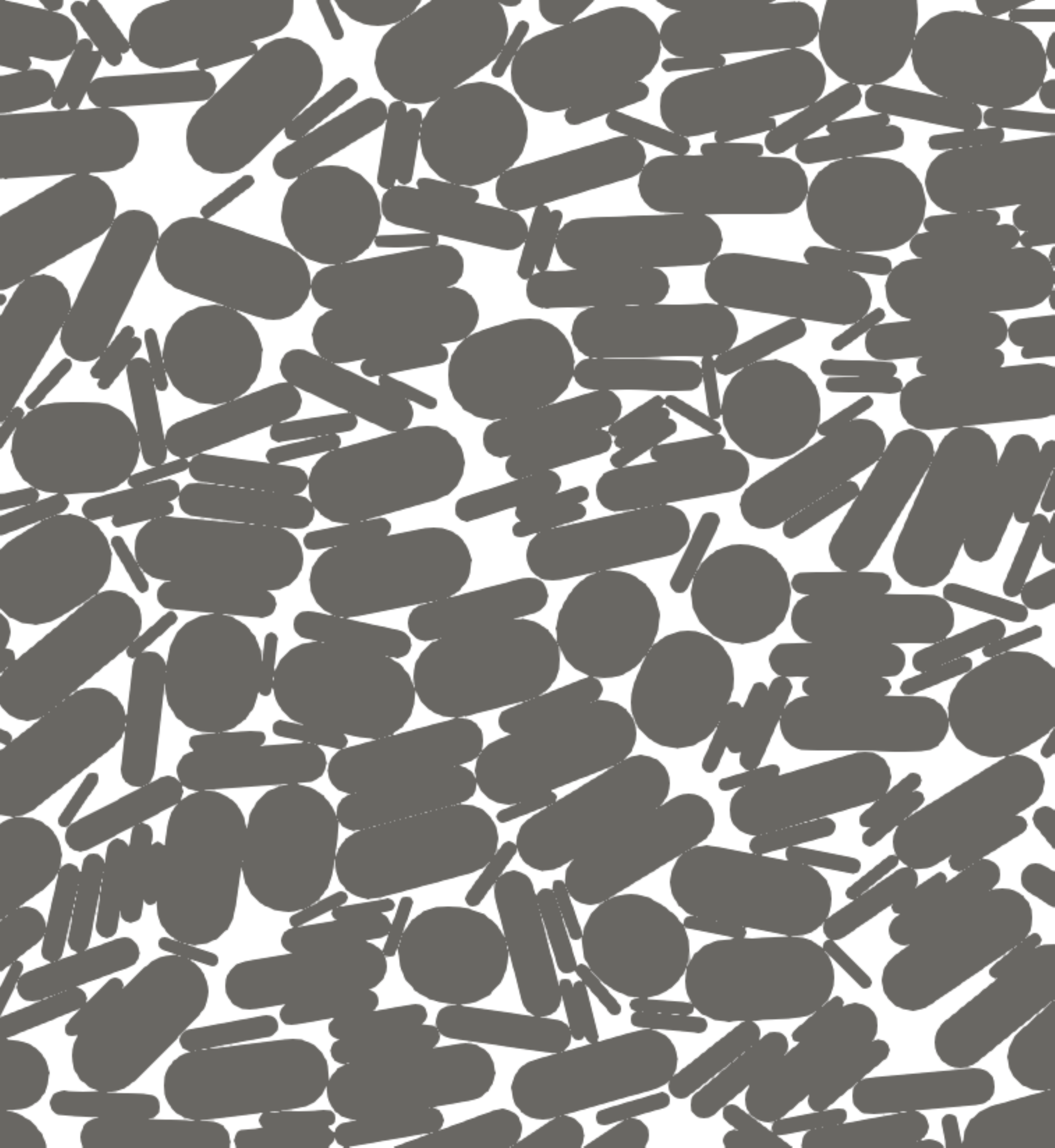}}
  \caption{Screenshots of samples at $\gamma = 400\%$ for case A and particle size dispersions $S = 0.4$(a), $S = 0.8$ (b). Figures (c) and (d) show the same particle size dispersions for case B.}
  \label{fig:ss_screenshot}
\end{figure}

We characterize the average configuration of the samples at the steady state by using information related to either the particles or their contacts. 
Elongated particles under shear deformation tend to rotate, and get aligned along the deformation orientation. 
Thus, they bear the load along their longest sides. 
To illustrate this, Fig. \ref{fig:ss_screenshot} shows some samples at $\gamma = 400\%$, in which we find most of the elongated particles pointing in the horizontal direction. 
For circular particles, the rotation is irrelevant. 
In turn, we observed that the alignment of elongated particles occurs gradually during shear but, remarkably, finds steady values for just slightly larger deformations than those seen for the stabilization of $\nu$ and $q/p$. 
In particular, we found that the average orientation of elongated grains (i.e., those with $\lambda > 1$) remains between $20^{\circ}$ to $30^{\circ}$ with respect to the horizontal for all spans $S$. 

\begin{figure}[tbh]
  \centering
  \includegraphics[width=0.45\linewidth]{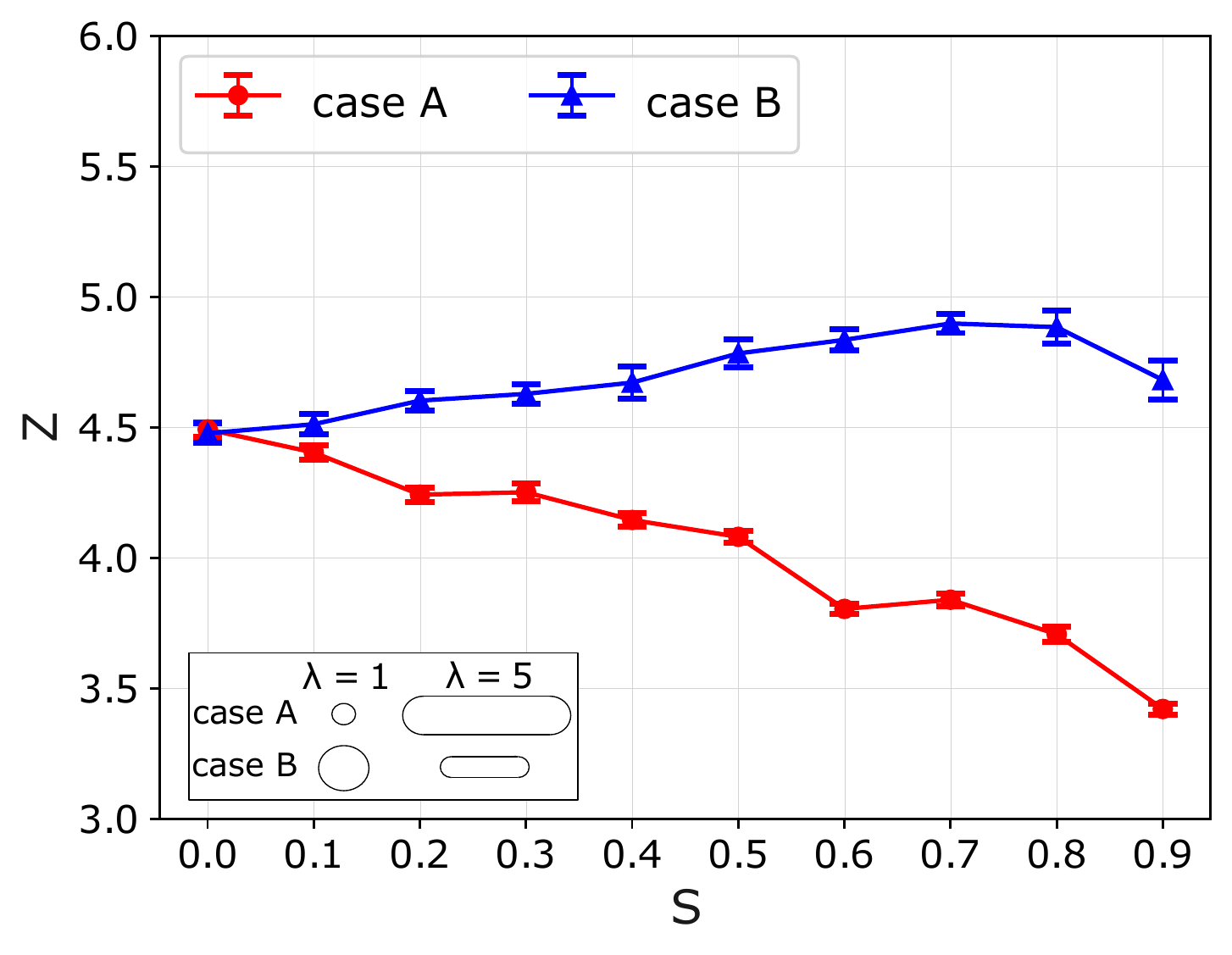}
  \caption{Evolution of the coordination number as a function of the particle size span for cases A and B.}
  \label{fig:coordination}
\end{figure}

On the side of the interactions between particles, we use as an indicator of their connectivity the average number of contacts per particle or \emph{coordination number}, defined as $Z = 2 N_c/N_p^*$, being $N_c$ the total number of force-bearing contacts, and $N_p^*$ the effective number of grains transmitting forces. 
Figure \ref{fig:coordination} shows $Z$ as a function of the particle size span $S$ for cases A and B. 
While for circular particles $Z$ is expected to remain close to $4$, independently of $S$ \cite{Roux2000}, the inclusion of elongated particles is known to affect the value of $Z$ at the steady state \cite{Azema2010,Boton2013}. 
As the particle size span increases, cases A and B present opposite evolutions with a decrease of $Z$ for assemblies containing big elongated particles (case B) and a drop of connectivity as the samples contain big circular particles (case A). 
This evolution of $Z$ is not evident in the screenshots presented above. 
Although for case A, the big elongated particles seem to gather many small particles around them, those increments of coordination turn out to be only local phenomena are not translated to the macroscale given the fewer large grains in the sample. 

\begin{figure}[htb]
  \centering
  \subfigure [ ] {\includegraphics[width=0.45\linewidth]{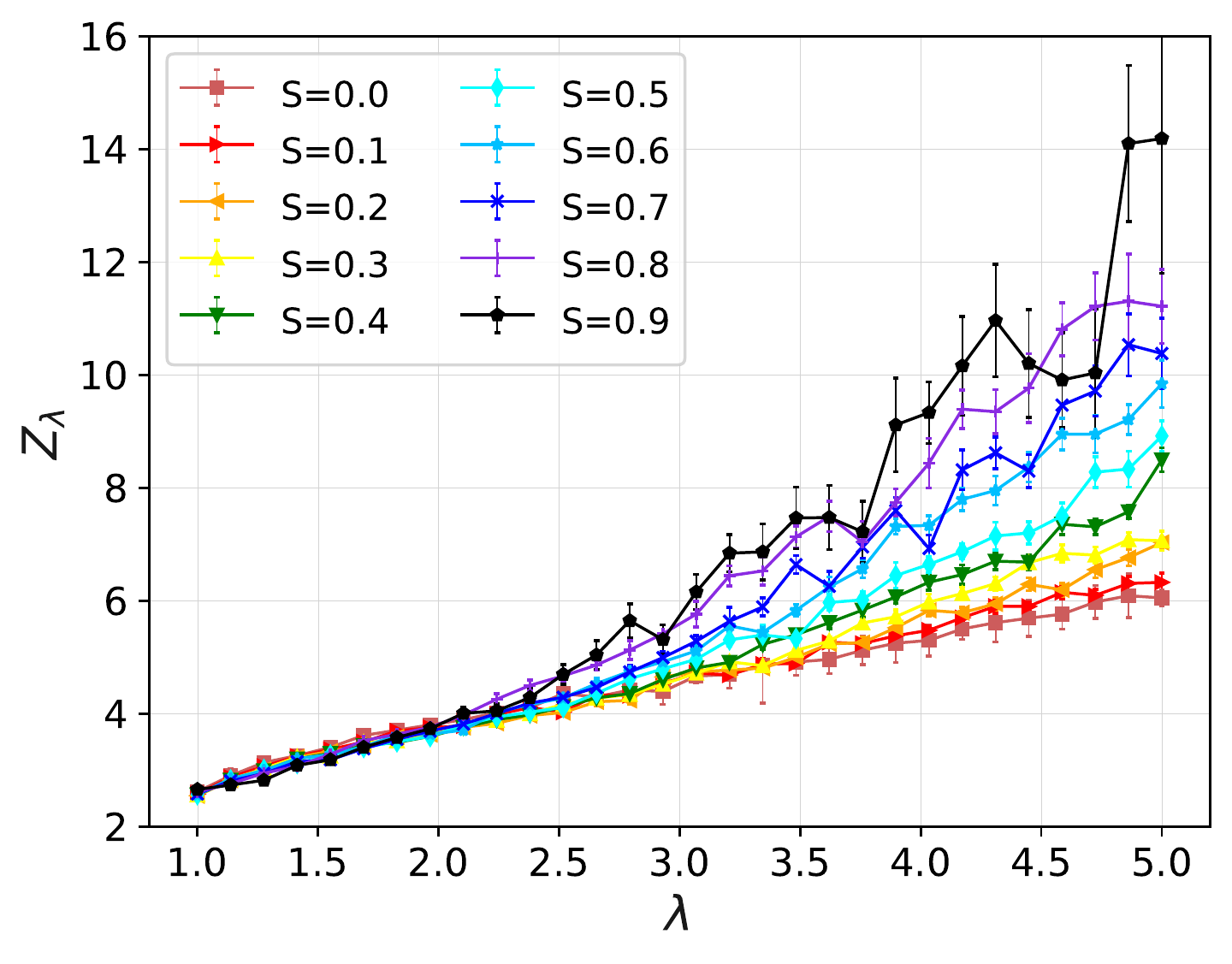}}
  \subfigure [ ] {\includegraphics[width=0.45\linewidth]{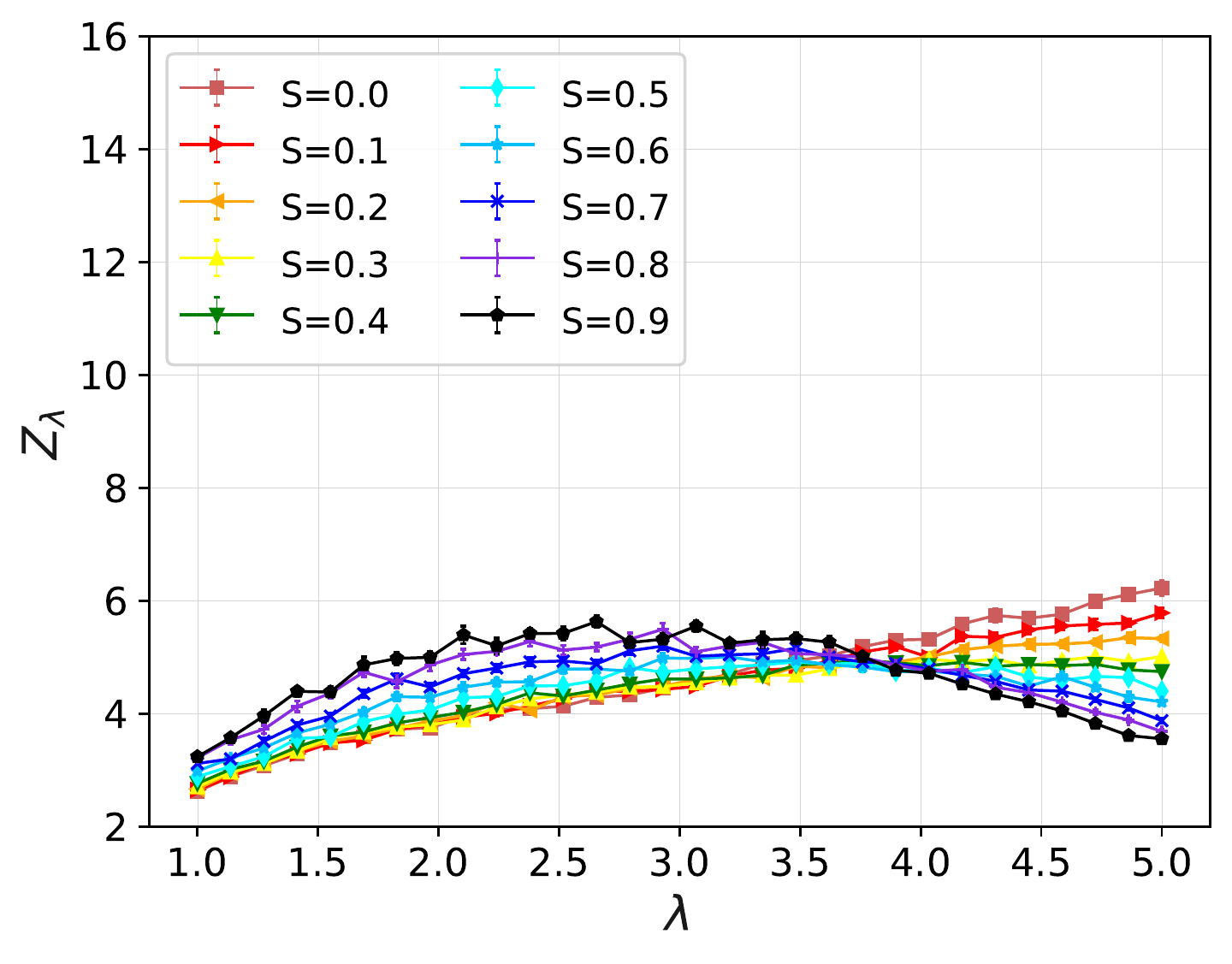}}
  \caption{Evolution of the average contact number by particle elongation $\lambda$ for case A (a) and case B (b). 
  Error bars present the standard deviation of the data.}
  \label{fig:coordination_shape}
\end{figure}

To understand these complex compensations of connectivity, we extract the coordination number by particle shape $Z_{\lambda}$ (i.e., the average number of contacts per particle of shape $\lambda$). 
Figure \ref{fig:coordination_shape}(a) presents the average number of contacts  $\lambda$ in case A. 
We see that $Z_{\lambda}$ increases with $\lambda$ in agreement with the observation that larger elongated grains are surrounded by many small circular grains with an average of $\simeq 13$ contacts per particle with elongation $\lambda=5$. 
The evolution of $Z_{\lambda}$ for case B is more complex (see Fig. \ref{fig:coordination_shape}(b)). 
As $\lambda$ increases, the increment of connectivity by particle shape changes from a linear trend for $S=0$, towards a parabolic shape for $S>0.6$, in which the larger circular grains are not the most connected particles in the assemblies. 
For those size disperse samples, the most connected shapes seem to have an aspect ratio around $\lambda \simeq 2.5$.
For that class of particles, the average number of contacts remains $Z_{\lambda=2.5} \simeq 5$. 

\begin{figure}[htb]
  \centering
  \subfigure [ ]{\includegraphics[width=0.33\linewidth]{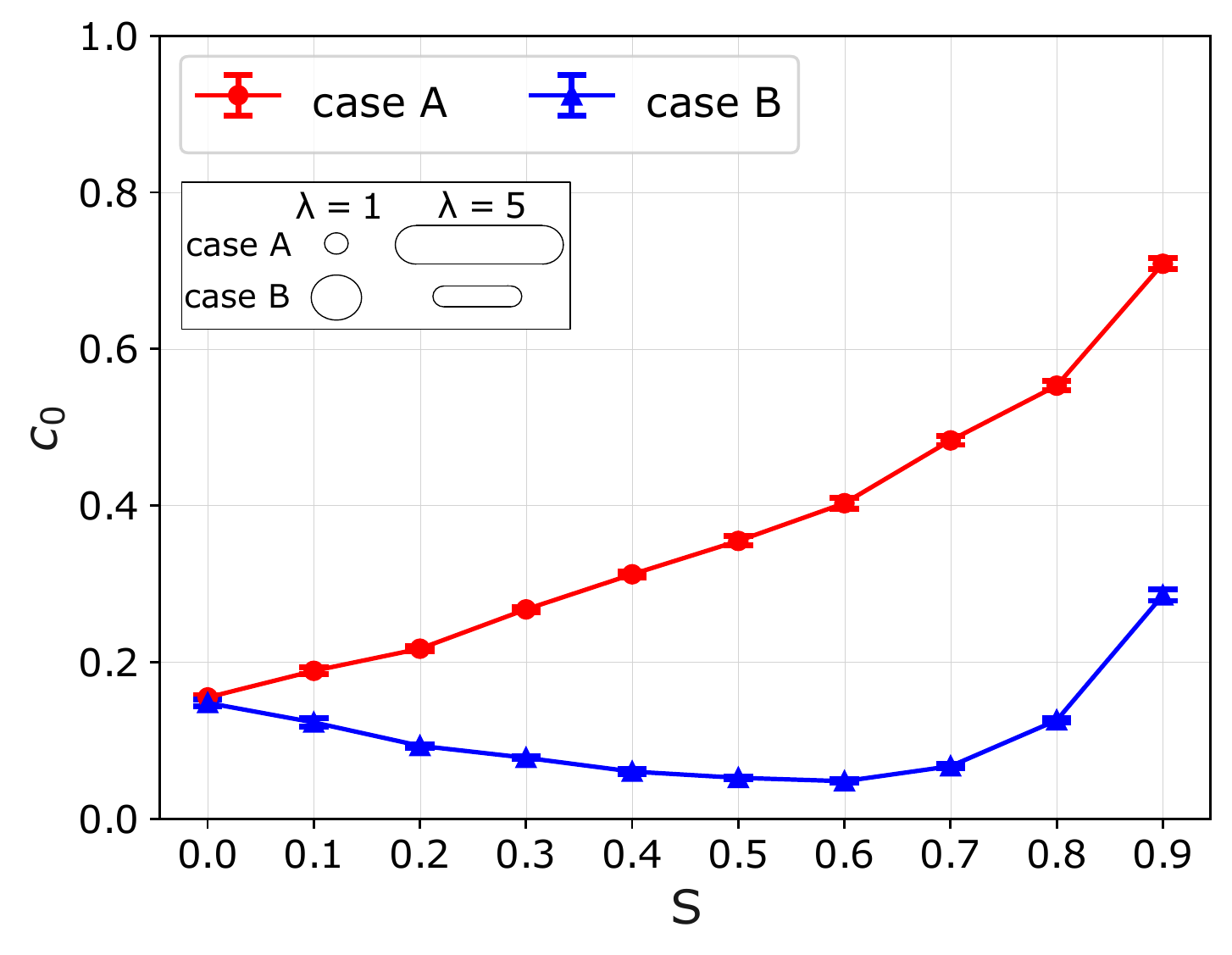}}
  \subfigure [ ]{\includegraphics[width=0.33\linewidth]{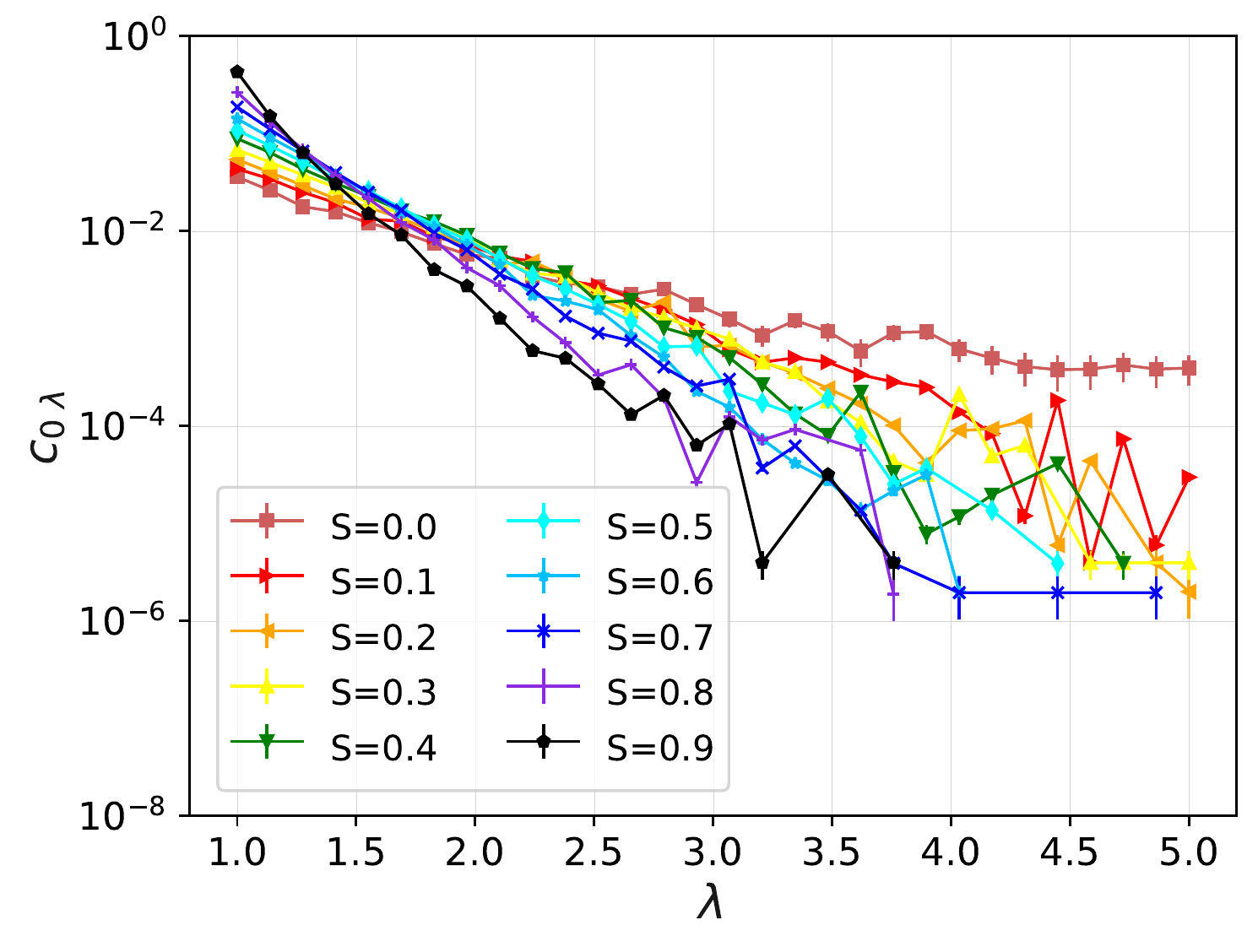}}
  \subfigure [ ]{\includegraphics[width=0.33\linewidth]{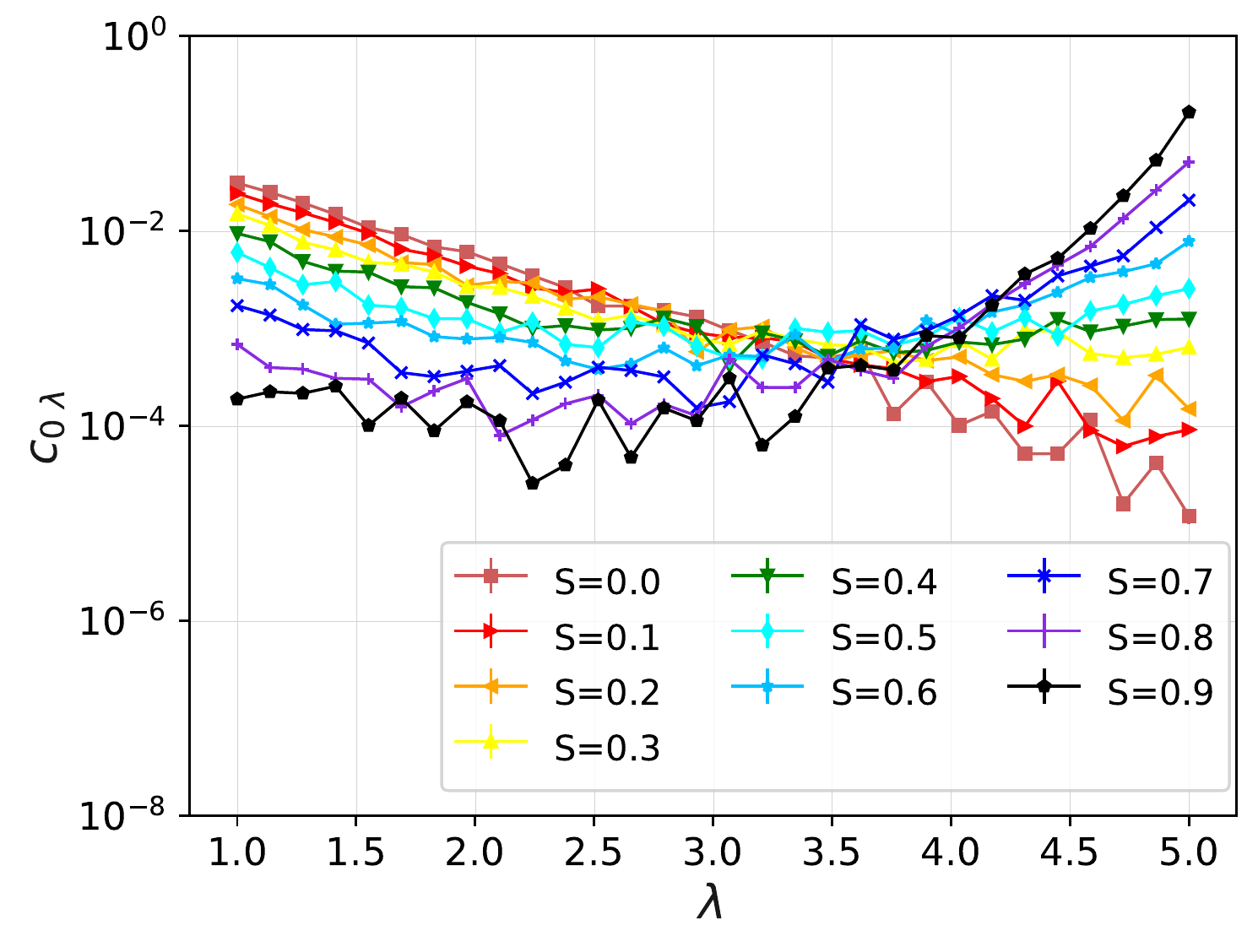}}
  \caption{(a) Evolution of the proportion of floating particles for cases A and B as a function of the particle size span $S$. Proportion of floating particles by shape class $\lambda$ for evolving particle size span $S$ and case A (b) and case B (c). 
  Error bars present the standard deviation of the data.}
  \label{fig:floating}
\end{figure}

Note that these parameters are computed over the particle number of a given shape class. 
However, the particle size distributions were set using uniform distributions by volume fractions. 
In order to identify the contribution of each shape class, it is important to quantify the number of particles excluded from this analysis since bodies having less than two active contacts (i.e., floating particles) are disregarded of the computation pf $N_p^*$. 
Figure \ref{fig:floating}(a) shows the evolution of the proportion of floating particles $c_0 = N_p^{0}/N_p$, with $N_p^0$ being the number of floating particles, as a function of $S$ for cases A and B. 
Case A shows $c_0$ gradually increasing with $S$ as reported in the literature \cite{Voivret2009, Nguyen2015, Cantor2018}. 
This phenomenon is supported by a large proportion of smaller particles rattling in the poral space (see Fig. \ref{fig:floating}(b) on the proportion of floating particles by shape class $\lambda$). 
Case B presents an unexpected nonlinear evolution of $c_0$ with $S$, first decreasing with $S$ up to $S=0.6$ and then rapidly increasing again. 
This is indeed related to the fact that small particles can fill the pores left between larger grains in which they are allowed to rattle. 
In terms of particle shape class participation, case B shows a compensation mechanism in which, for low values of $S$, the floating particles mostly belong to the class of smaller grains (see Fig. \ref{fig:floating}(c)). 
However, as $S$ increases, the small class of particles becomes more active at the load transmission while the bigger particles become less active and rattle.

\begin{figure}[htb]
  \centering
  \subfigure {\includegraphics[width=0.25\linewidth]{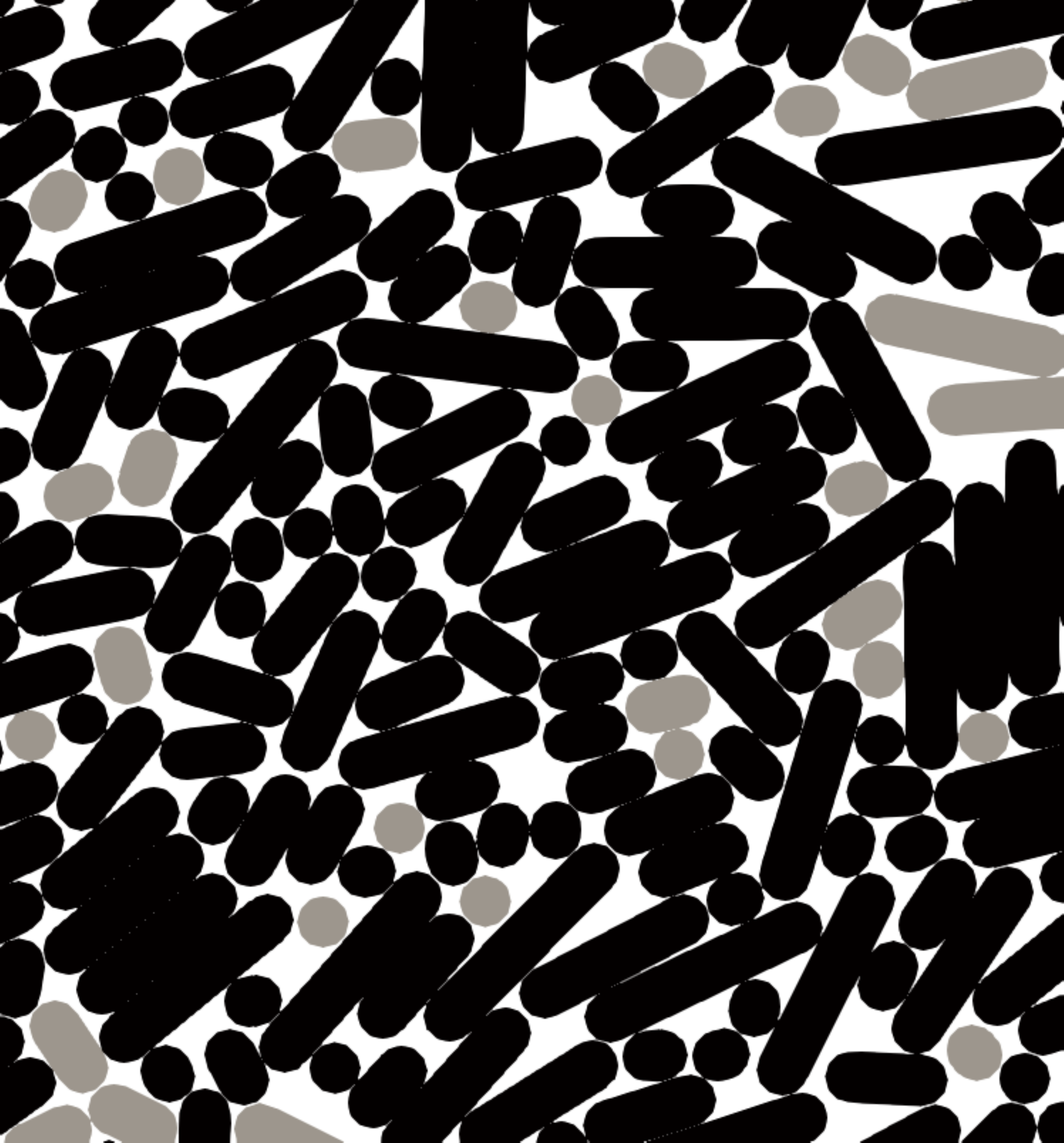}}
  \subfigure {\includegraphics[width=0.25\linewidth]{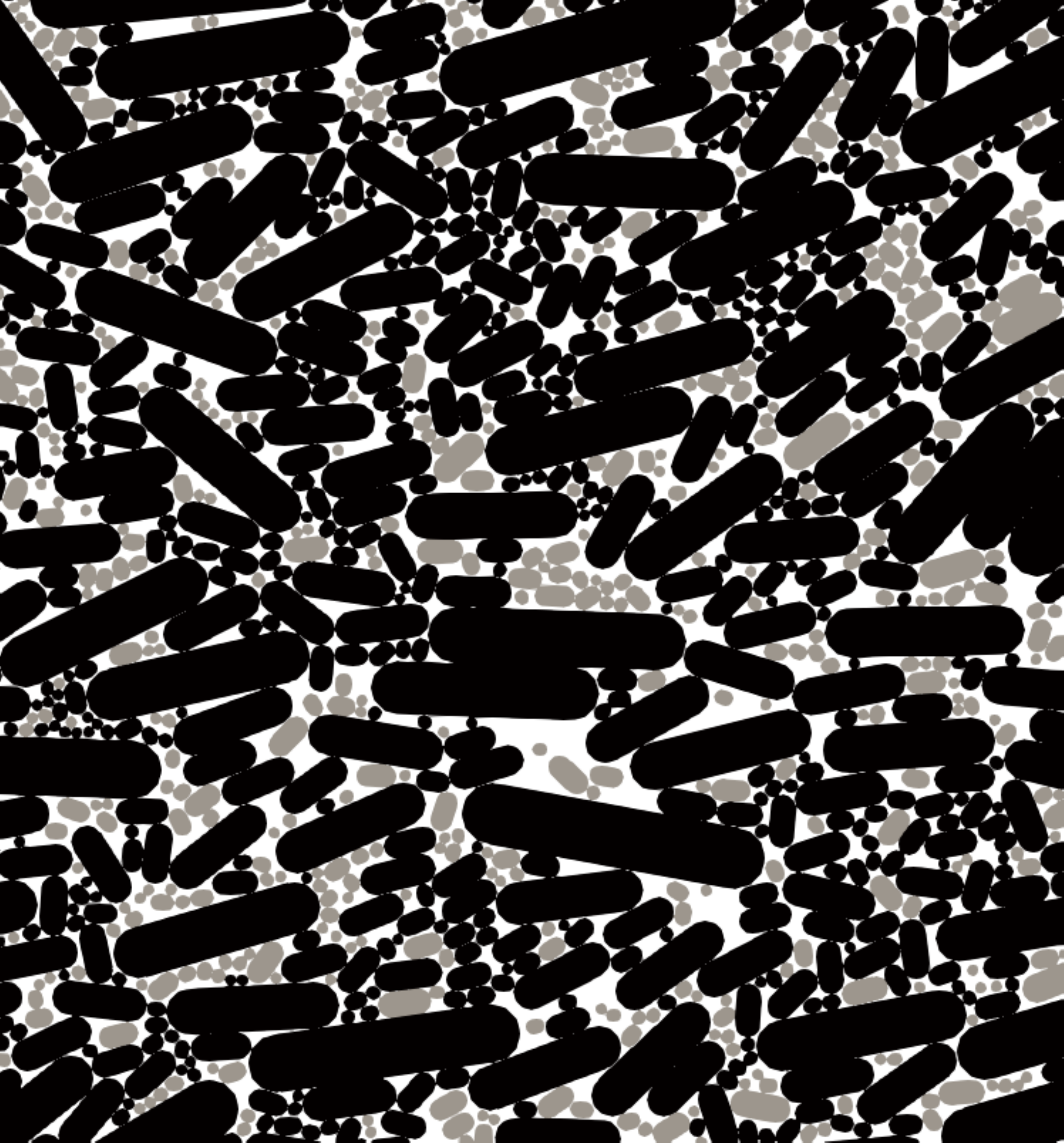}}
  \subfigure {\includegraphics[width=0.25\linewidth]{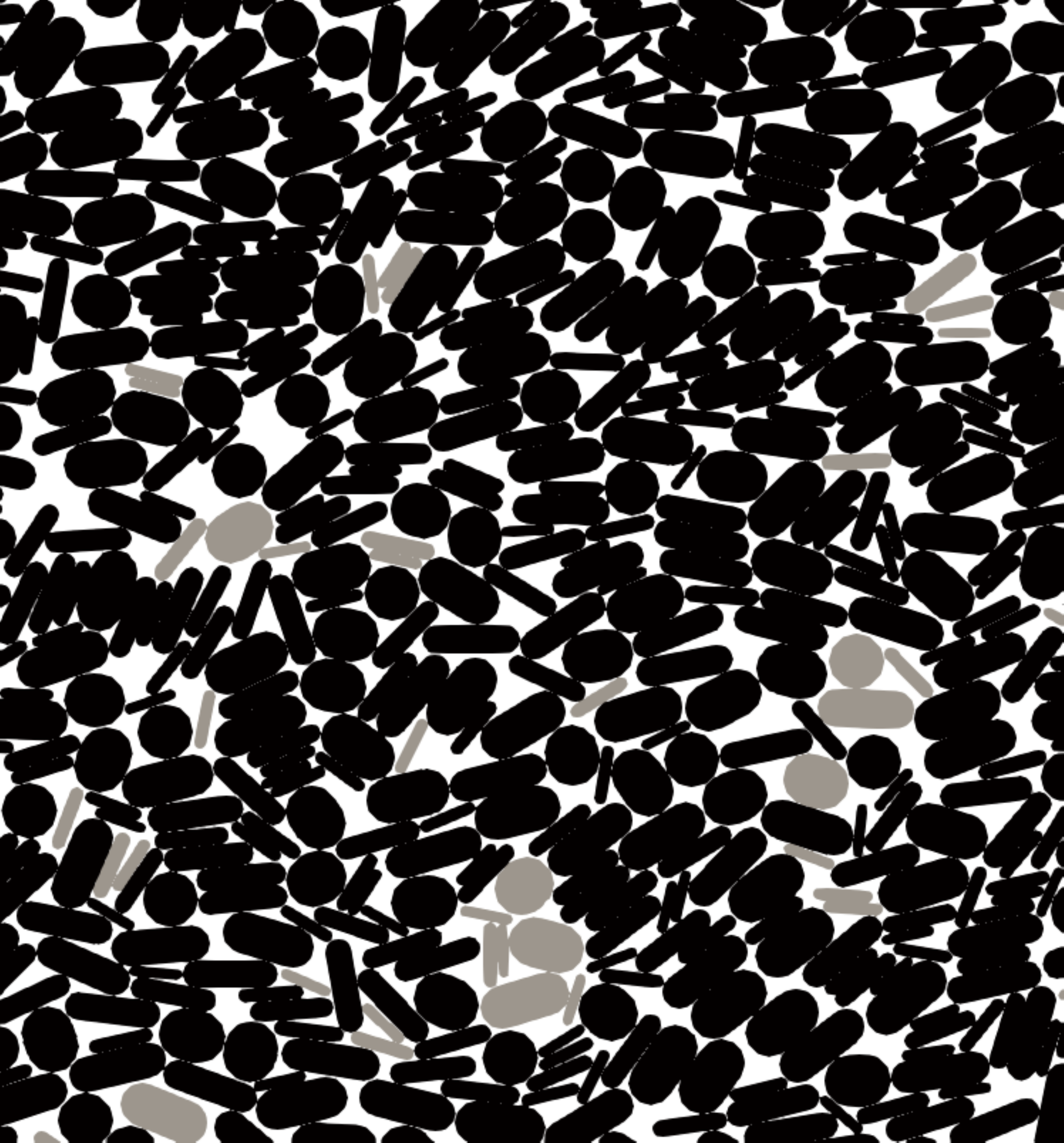}}
  \subfigure {\includegraphics[width=0.25\linewidth]{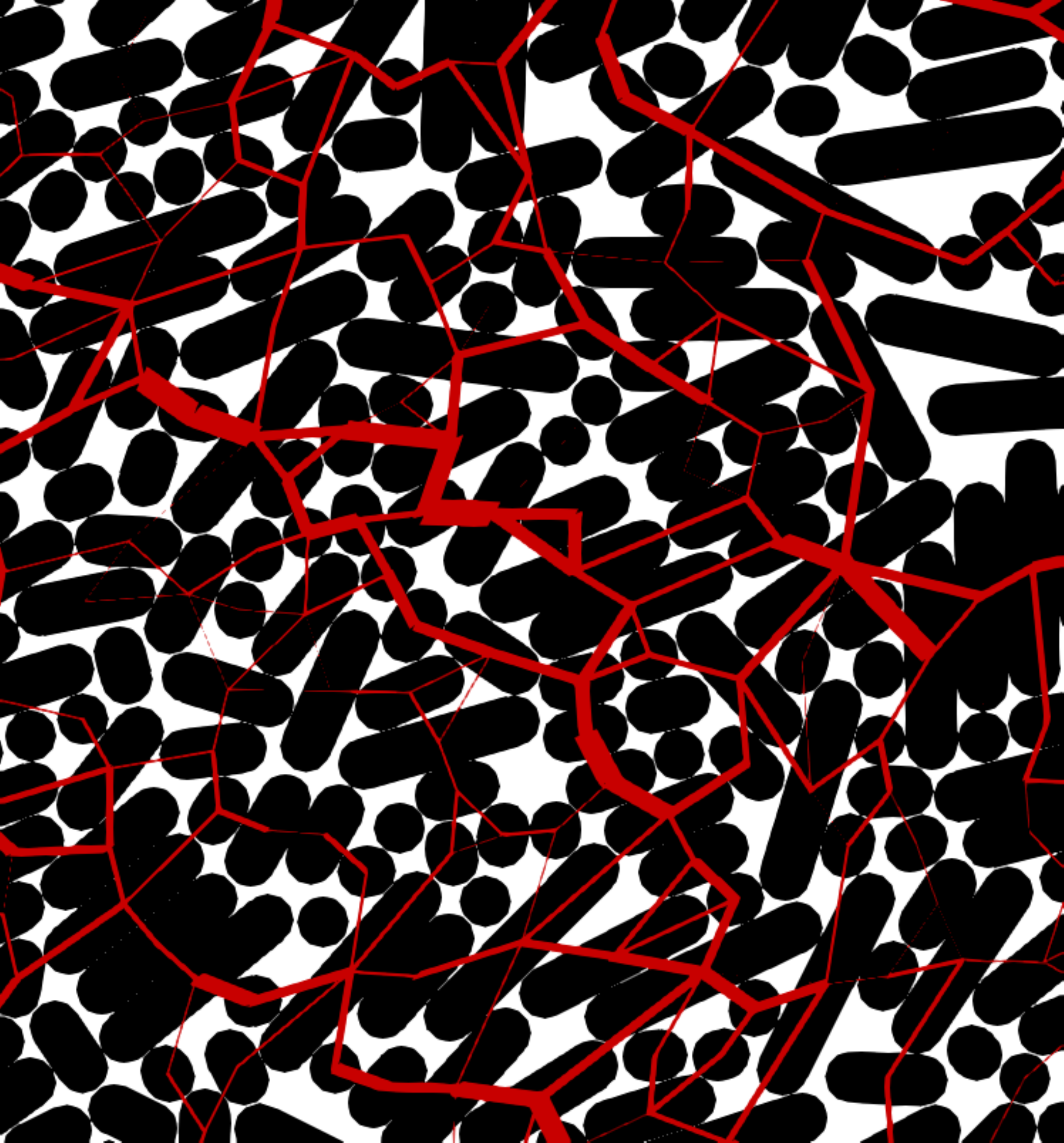}}
  \subfigure {\includegraphics[width=0.25\linewidth]{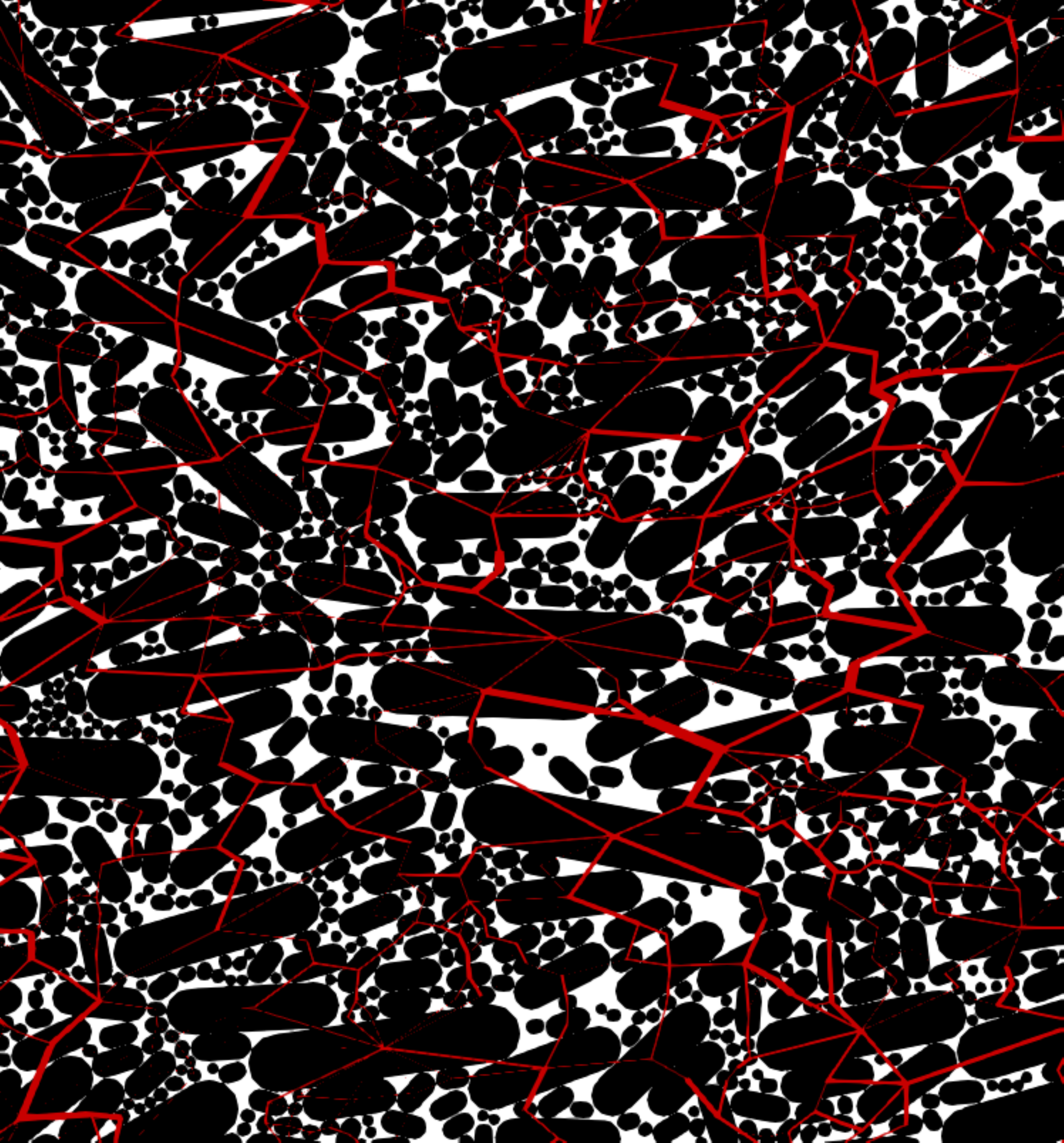}}
  \subfigure {\includegraphics[width=0.25\linewidth]{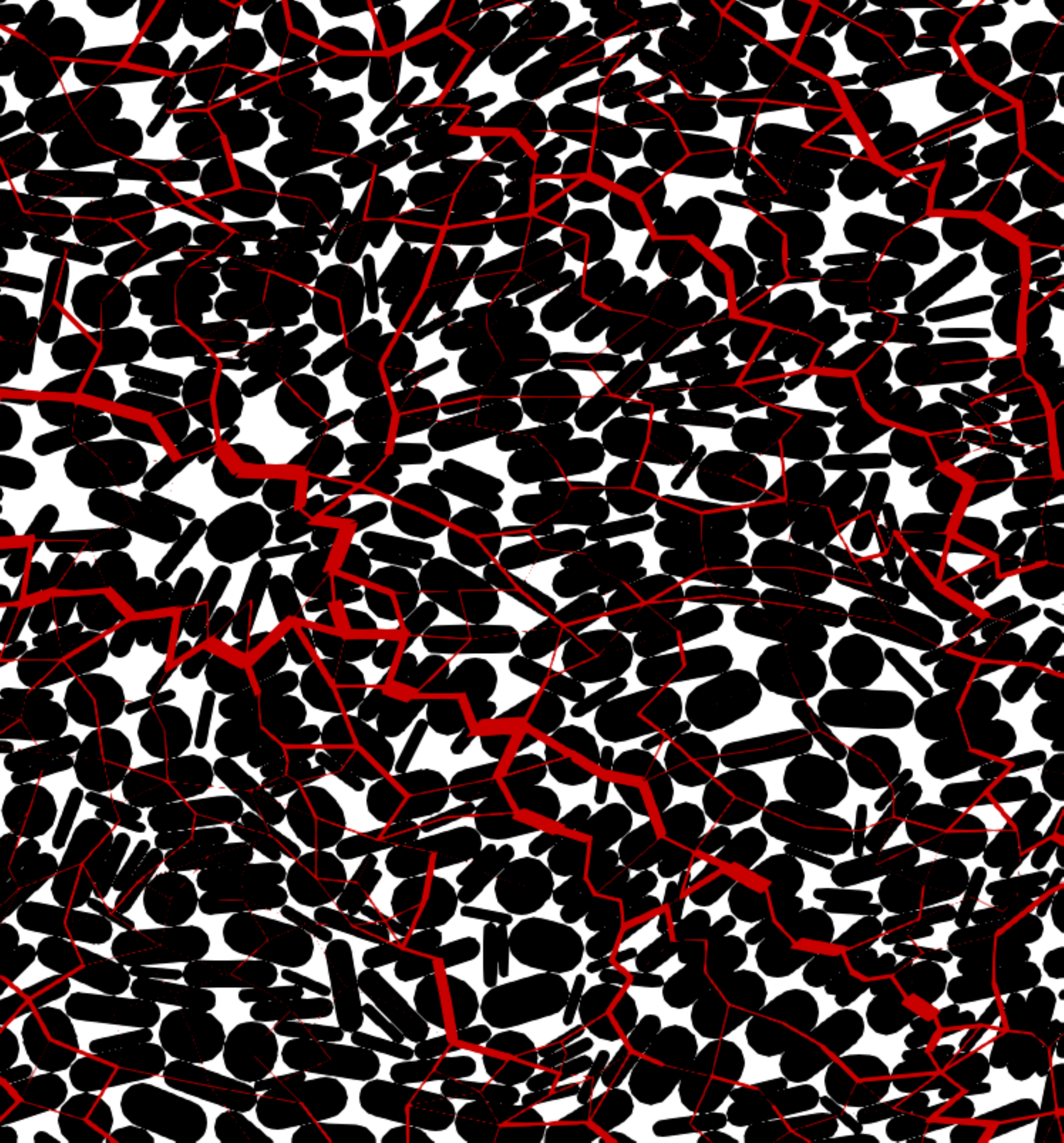}}
  \caption{(Top) Screenshots highlighting floating particles (gray) and active particles (black) for samples with $S=0$ (left), and $S=0.7$ for case A (middle) and case B (right). (Bottom) Force networks for the same set of samples.}
  \label{fig:floating_forces}
\end{figure}

In order to illustrate the evolution of the floating particle proportions, Fig. \ref{fig:floating_forces} (top) shows screenshots of the samples in which floating particles are shown in black and active particles in gray for different values of $S$ for cases A and B. 
At the bottom of the same figure, we present the force transmission within the same samples as bars connecting the center of mass of the particles. 
The thickness of the bars is proportional to the force intensity at the contacts. 
We clearly observe the different roles bigger particles have at the force transmission in these samples. 
While in case A, most of the load is transmitted through the large particles, in case B, more granulometric classes - and in effect, more shape classes - are involved at the force transmission resting importance to the bigger grains. 

\section{Micromechanical contributions to strength}\label{sec:roles}
To understand the drop of strength for case A as a function of $S$ and the apparent independence of strength on $S$ for case B, it is useful to use the decomposition of the granular stress tensor in terms of micromechanical descriptors as proposed by \cite{Rothenburg1989}.

\begin{figure}[htb]
  \centering
  \includegraphics[width=0.6\linewidth]{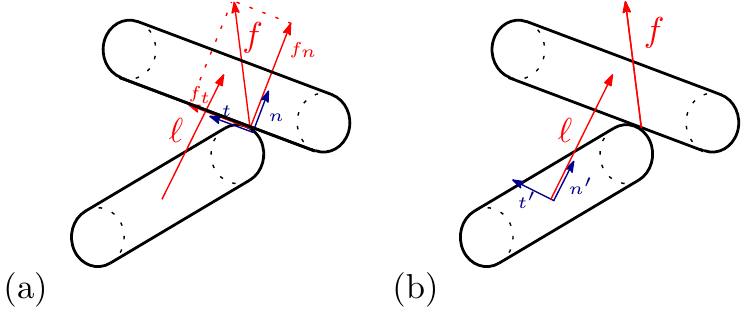}
  \caption{Scheme of (a) the reference frame created by contacts between particles, and (b) the reference frame created by branch vectors.}
  \label{fig:schemes}
\end{figure}

This approach is based on the fact that contacts, forces, and branch vectors have periodic distributions in space. 
These distributions can be computed in two different forms. 
Any interaction between two particles defines a frame having a unitary normal and tangential orientations $\bm{n}$ and $\bm{t}$, that allows one to express the contact force as $\bm{f} = f_n \bm{n} + f_t \bm{t}$, with $f_n$ and $f_t$ the normal and tangential contact forces, respectively. 
On this same frame, the branch vector can be written as $\bm{\ell} = \ell_n \bm{n} + \ell_t \bm{t}$, being $\ell_n$ and $\ell_t$ the projections of $\ell$ on the contact frame. 
Frame $\{\bm{n}, \bm{t}\}$ has been used in many studies to analyze the microstructural properties of regular monosize granular materials. 
However, for elongated grains, another reference frame has proven to be more adequate \cite{Azema2010,Cantor2021}. 
The branch vector can alternatively define a frame with $\bm{n}'$ the unitary vector in the direction of $\bm{\ell}$, and $\bm{t}'$ a tangential orientation defined by a $90^{\circ}$ counterclockwise rotation of $\bm{n}'$ (see Fig. \ref{fig:schemes}).
Using vector $\bm{n}' = \{n'_x, n'_y\}$, we define the orientation of the branches as $\theta =  \mathrm{acos}(n'_x)$. 

\begin{figure}[htb]
  \centering
  \subfigure {
  \includegraphics[width=0.38\linewidth]{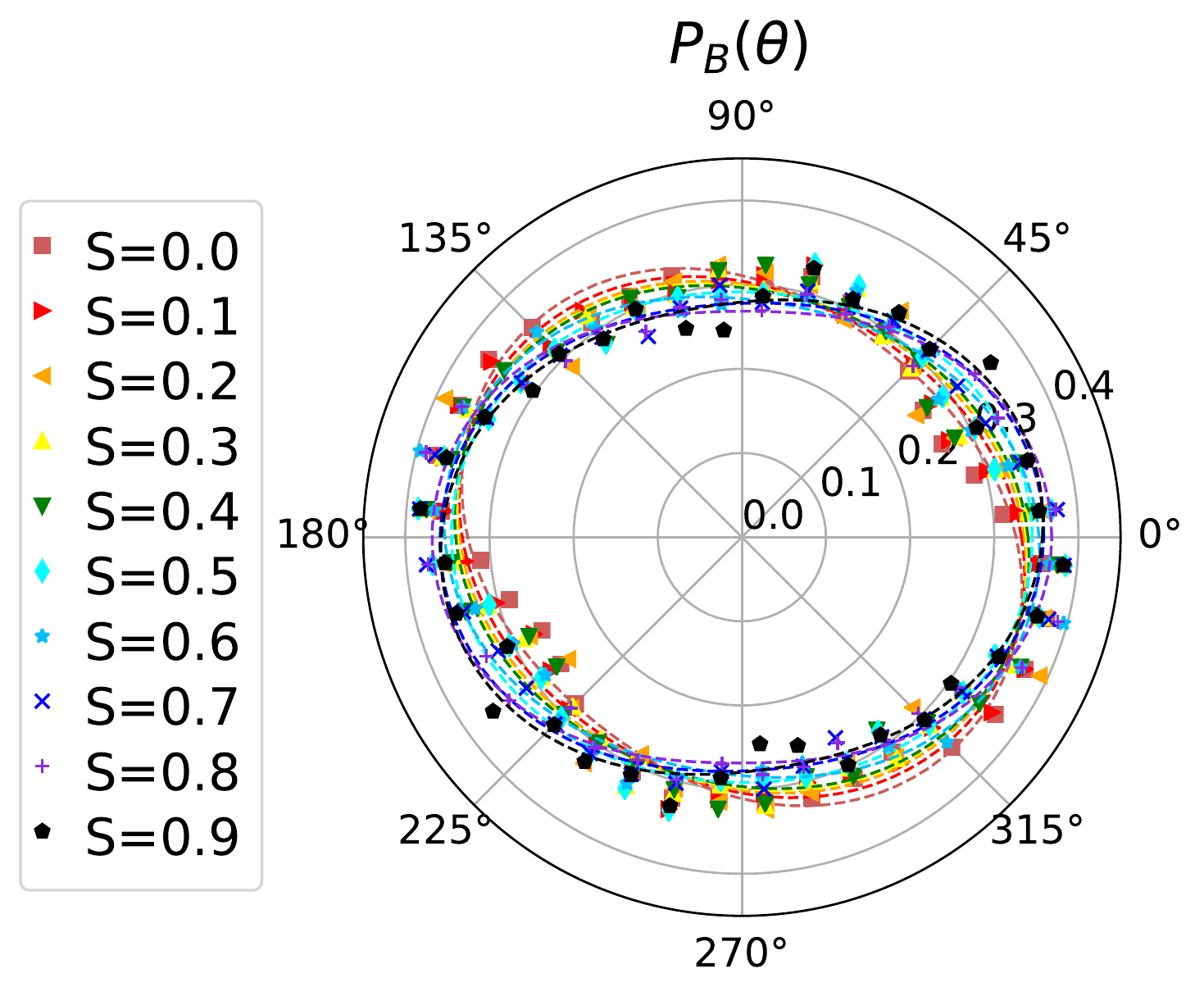}}
  \subfigure {
  \includegraphics[width=0.3\linewidth]{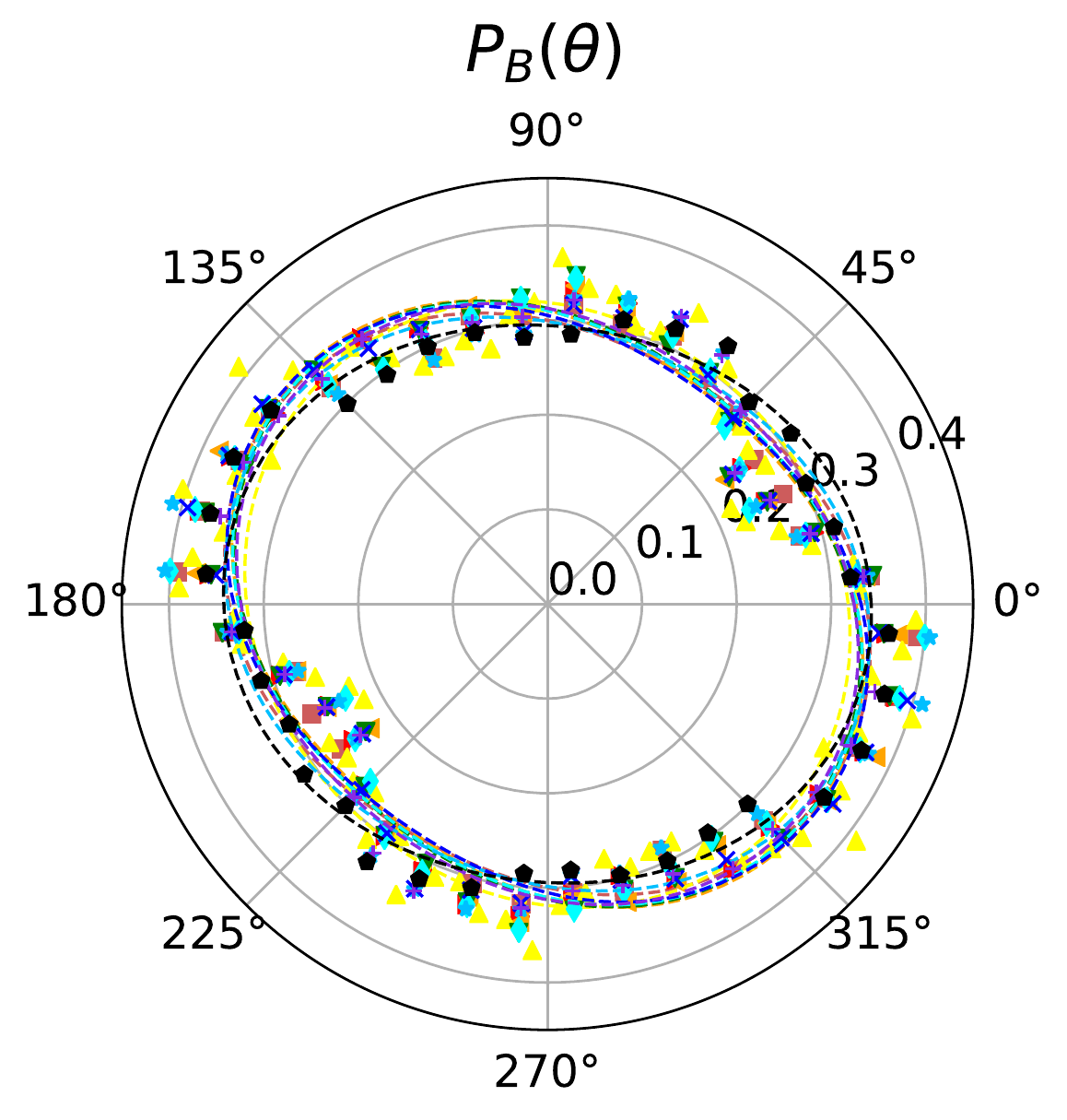}}
  \caption{Angular distribution of branch orientations for cases A and B and different values of size dispersion $S$.}
  \label{fig:pc_theta}
\end{figure}

Standing on the branch frame, we can define the branch orientation probability as 
\begin{equation}
P_{B}(\theta) = N_B(\theta)/N_B, 
\end{equation}
where $N_B(\theta)$ is the number of branches pointing at angle $\theta$, and $N_B$ is the total number of active branches in the sample (i.e., branches linked to two grains that carry a non-zero force at the contact). 
Figure \ref{fig:pc_theta} presents the distribution $P_{B}$ for cases A and B and different values of $S$. 
These angular distributions can be described using truncated Fourier series as 
\begin{equation}\label{eq:fourier_pb}
P_{B}(\theta) = \frac{1}{2\pi} \{1 + a_{B} \cos 2(\theta - \theta_{B})\}, 
\end{equation}
with $\theta_{B}$ the preferential orientation of the distribution and $a_{B}$ the level of anisotropy. 
Simple fitting of this equation to our data allows us to find these two parameters of the distribution. 
Note that in case A, the distribution $P_B(\theta)$ is roughly ellipsoidal with an orientation of the major axis varying from $\theta_{B} \simeq 135^{\circ}$ for $S=0$ to $\theta_{B} \simeq 180^{\circ}$ for $S=0.9$. 
In turn, for case B, the main orientation of the distributions fluctuates but remains close to $\theta_{B} \simeq 135^{\circ}$. 
In the same figures, we plot Eq. (\ref{eq:fourier_pb}) for each value of $S$ using dashed lines. 
We remark that the fit is not perfectly matching the discrete distribution of $P_B(\theta)$ for all orientations. 
In particular, for perpendicular orientations with respect to $\theta_B$, Eq. \ref{eq:fourier_pb} seems to underestimate the value for $P_B$.
When using regular or mono size grains, these first-order Fourier series have proven to capture very well the angular distributions. 
However, the introduction of elongated grains produces deep perturbations in the microstructure and geometrical organization of the particles. 

For the analysis of force transmission, note that the total vector force $\bm{f}$ can be projected on the branch frame letting us find a force acting between the centers of mass of touching grains called \emph{radial force} $f'_n$, and a shearing equivalent force along with the direction of $\bm{t}'$, called \emph{ortho-radial force} $f'_t$. 
For the radial forces, we can compute their angular distribution as 
\begin{equation}
\langle f_n^{'} \rangle(\theta) = \frac{1}{N_B(\theta)} \sum_{B \in \theta} f_n^{'},
\end{equation}
with $\langle \dots \rangle$ being the average radial force, and $N_B(\theta)$ the number of branches pointing at angle $\theta$. 
Similarly, for the ortho-radial forces, the angular distribution can be found as 
\begin{equation}
\langle f_t^{'} \rangle(\theta) = \frac{1}{N_B(\theta)} \sum_{B \in \theta} f_t^{'}.
\end{equation}

\begin{figure}[htb]
  \centering
  \subfigure [ ] {\includegraphics[width=0.38\linewidth]{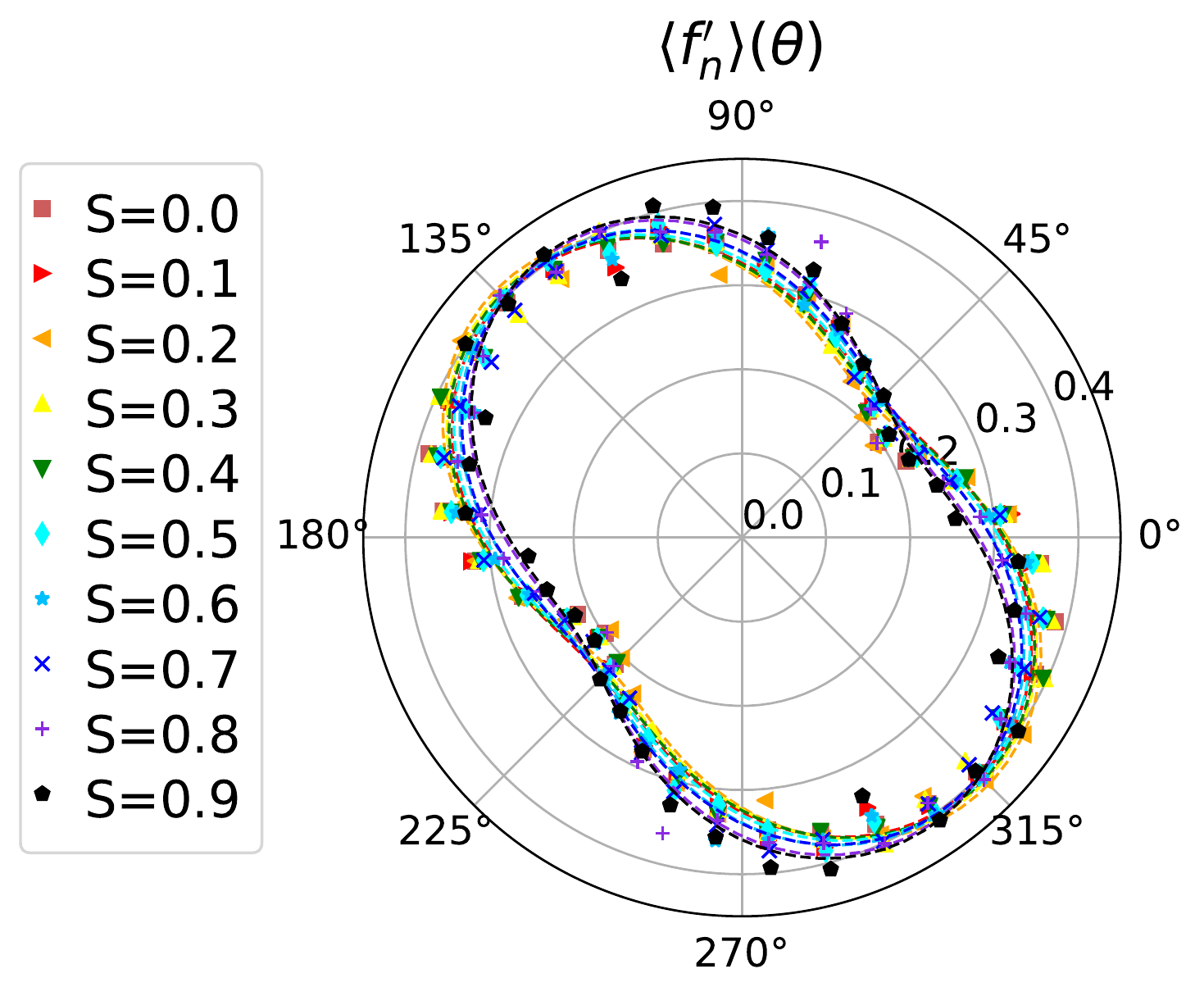}}
  \subfigure [ ] {\includegraphics[width=0.3\linewidth]{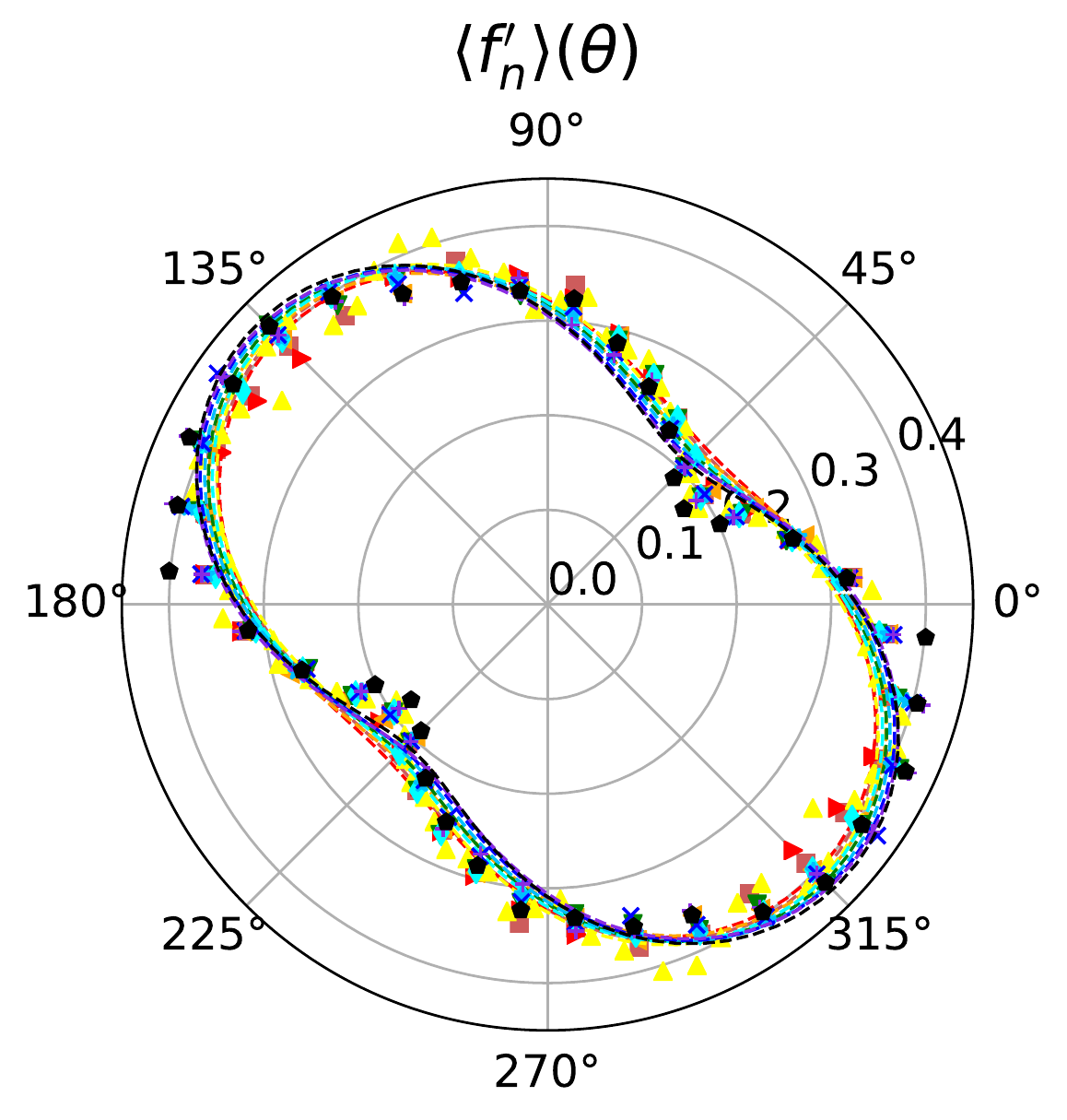}}
  \subfigure [ ] {\includegraphics[width=0.38\linewidth]{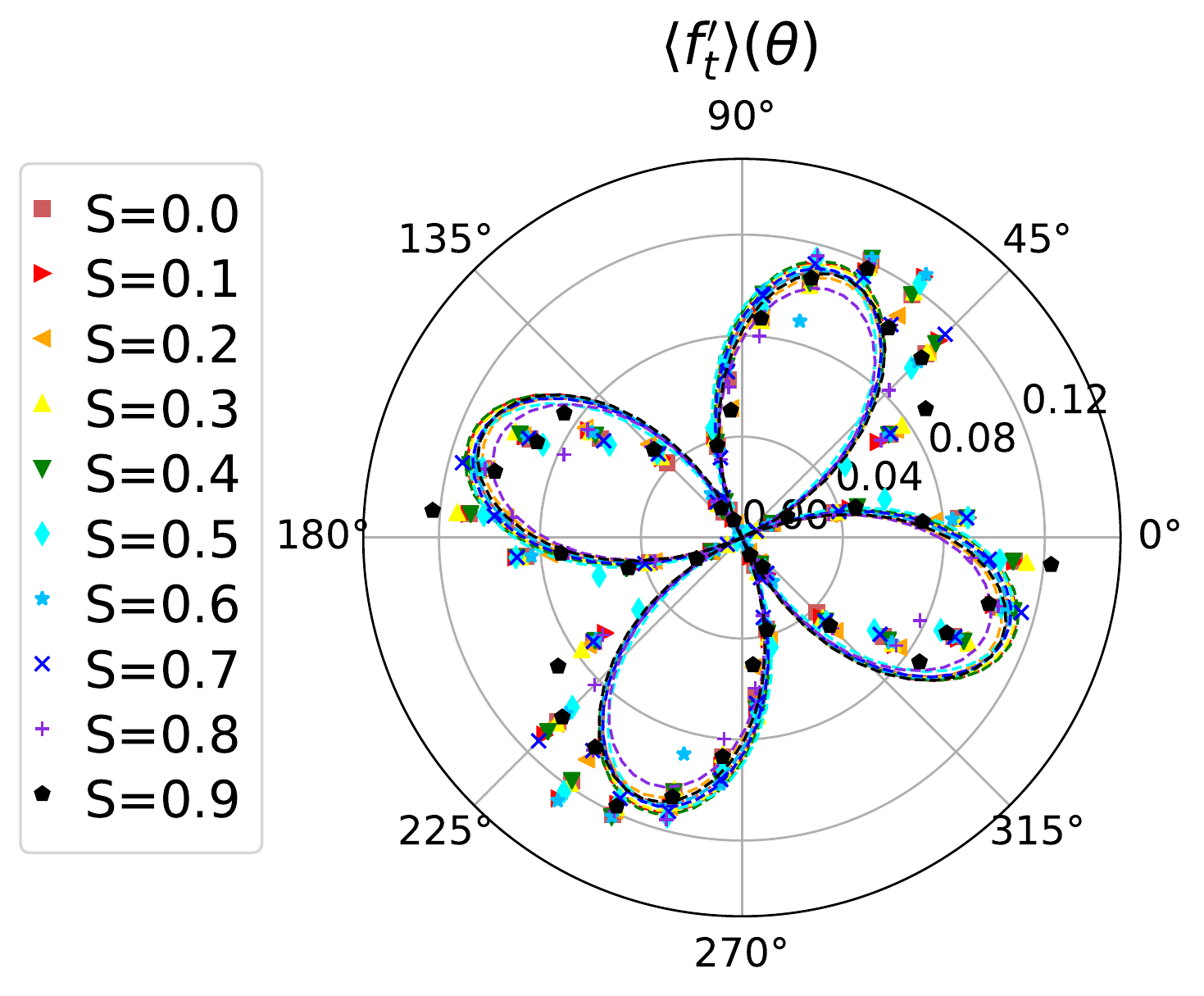}}
  \subfigure [ ] {\includegraphics[width=0.3\linewidth]{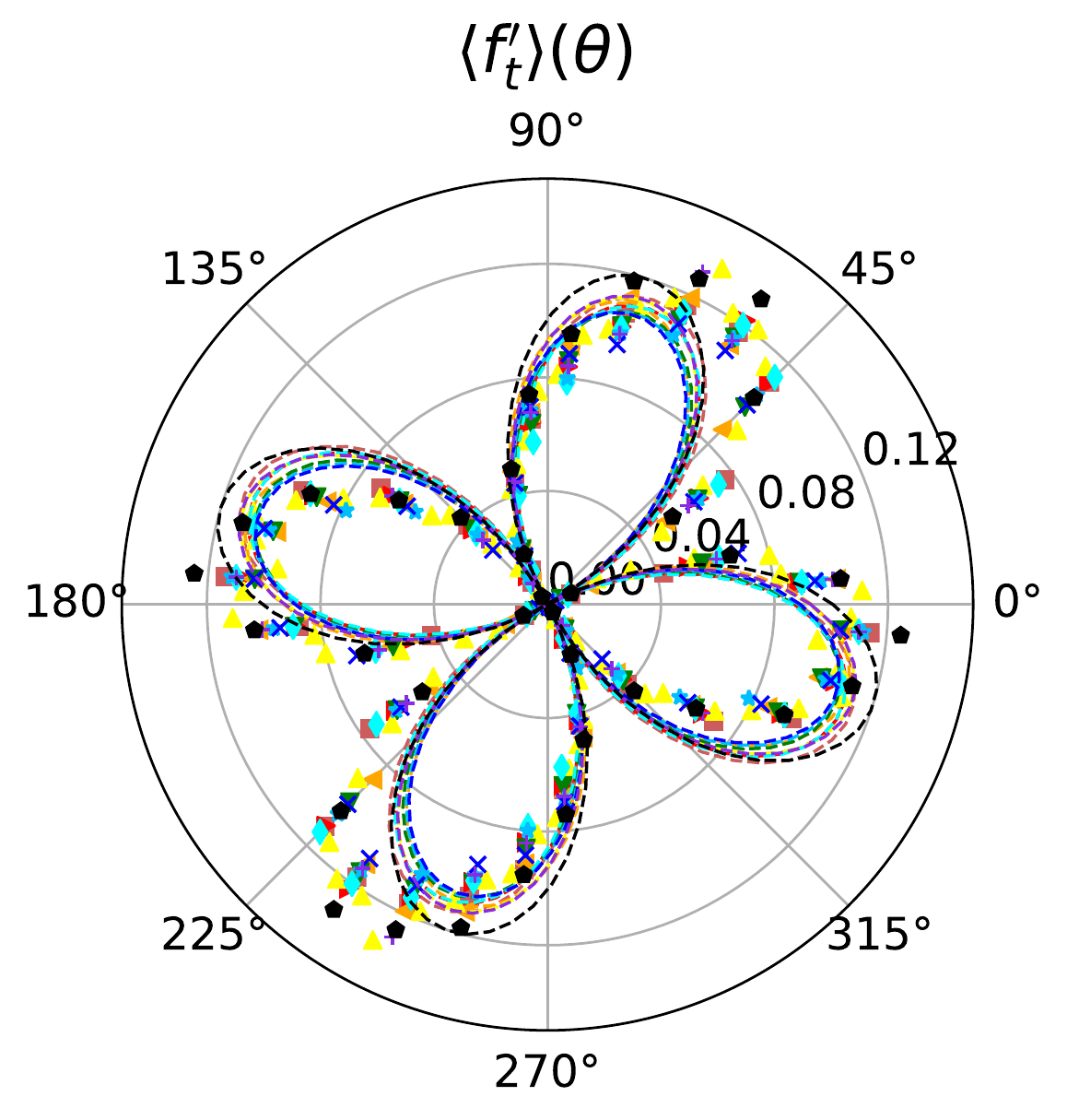}}
  \caption{Angular distribution of radial and ortho-radial force intensities for cases A and B and different grain size span $S$.}
  \label{fig:f_theta}
\end{figure}

Figure \ref{fig:f_theta} presents these angular distributions for cases A and B and different values of $S$. 

These angular force distributions can be described using truncated Fourier series as 
\begin{equation}
\langle f'_n \rangle(\theta) = \langle f'_n \rangle \{1 + a_{f_n} \cos 2(\theta - \theta_{f_n})\},
\end{equation}
and 
\begin{equation}
\langle f'_t \rangle(\theta) = -\langle f'_n \rangle \{a_{f_t} \sin 2(\theta - \theta_{f_t})\},
\end{equation}
with $\langle f'_n \rangle$ the average radial force in the sample, and $a_{f_n}$ and $a_{f_t}$ the radial and ortho-radial force anisotropies, and $\theta_{f_n}$ and $\theta_{f_t}$ the preferential orientations of each distribution. 
We observe that for both cases A and B, these radial force distributions are relatively similar, having larger force intensities at orientations $\theta \simeq 135^{\circ}$. 
In addition, the distributions seem independent of grain size span $S$. 
The ortho-radial force distributions have a four-fold symmetry that is roughly $\pi/2$ periodic. 
This periodicity is strong among rounded mono-size particles. 
Nevertheless, as we can observe in Figs. \ref{fig:f_theta}(c) and (d), elongated particles slightly break this symmetry, so a small mismatch can be observed between the discrete distribution and the Fourier functional forms trying to fit them. 

\begin{figure}[htb]
  \centering
  \subfigure [ ] {\includegraphics[width=0.38\linewidth]{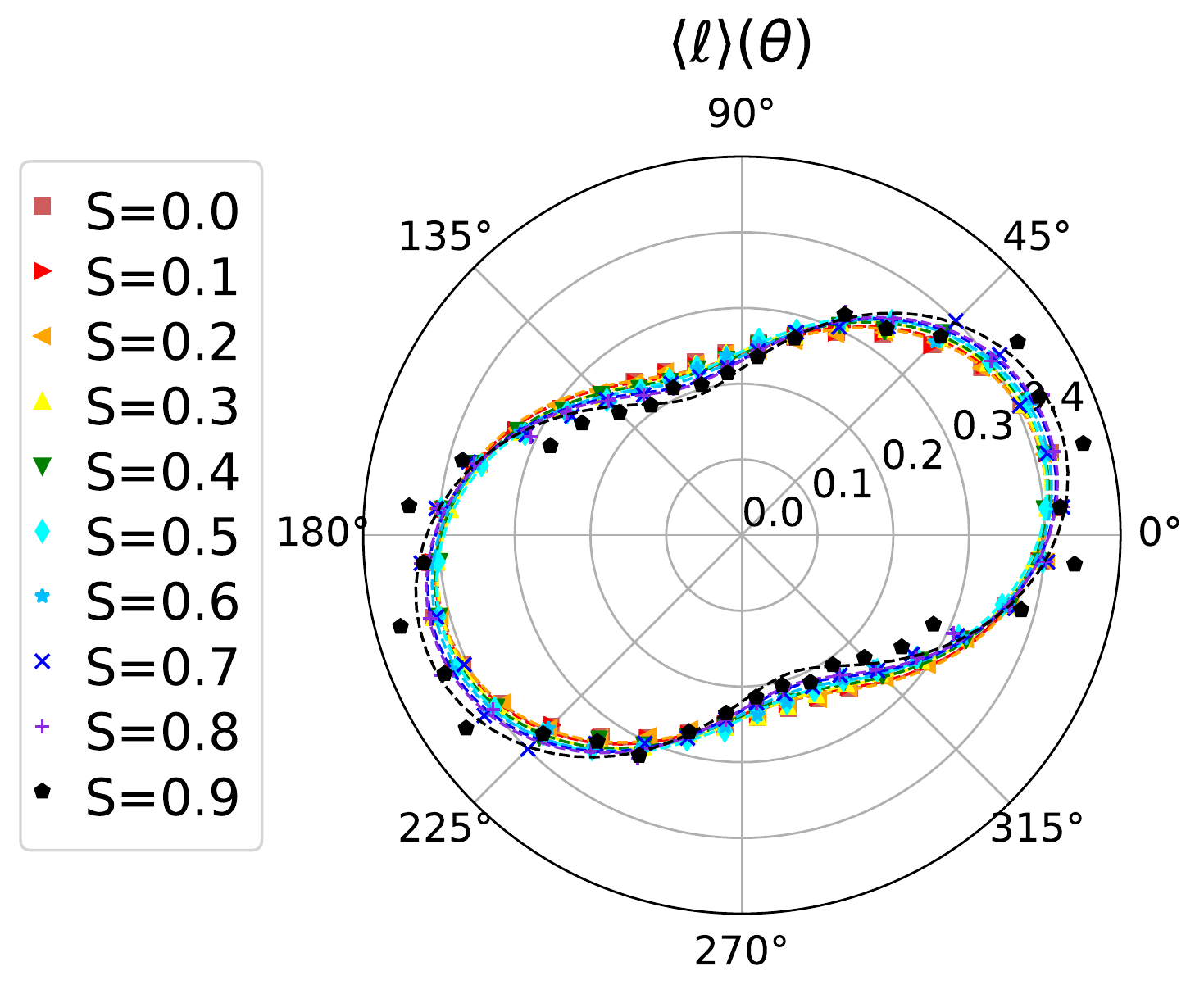}}
  \subfigure [ ] {\includegraphics[width=0.3\linewidth]{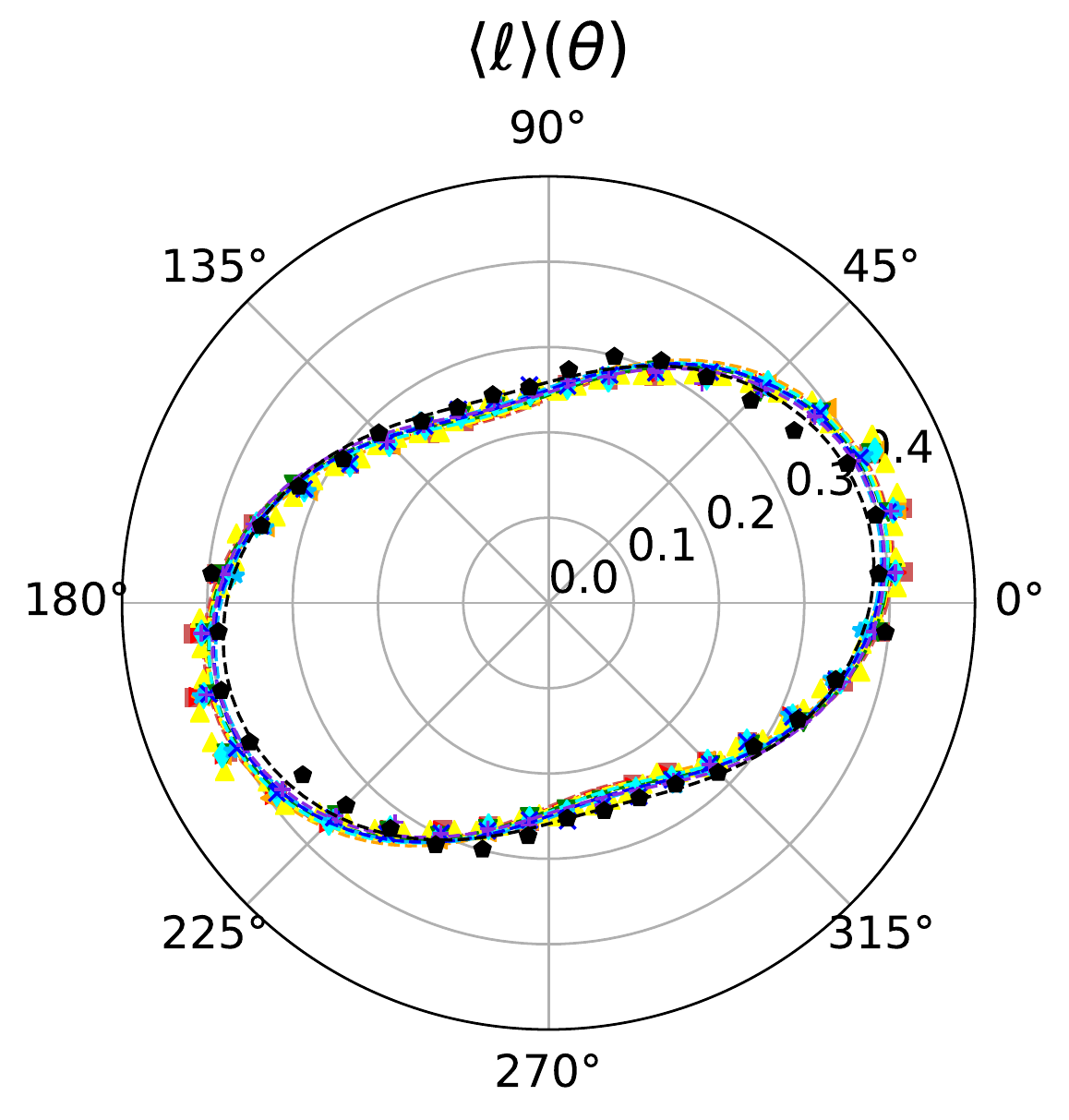}}
  \caption{Angular distributions of branch lengths cases A (a) and B (b).}
  \label{fig:l_theta}
\end{figure}

In addition, the analysis of the branch length is more simple since they can be written as $\bm{\ell} = \ell \bm{n}'$.
Under this definition, the branch vector does not have a tangential component, so only one level of anisotropy is related to this distribution. 
Figure \ref{fig:l_theta} presents the angular distribution of branch lengths for cases A and B and varying particle size span $S$. 

These angular branch distributions can be described as 
\begin{equation}
\langle \ell \rangle(\theta) = \langle \ell \rangle \{1 + a_{\ell} \cos 2(\theta - \theta_{\ell})\}, 
\end{equation}
with $\langle \ell \rangle$ being the average branch length in the sample, $a_{\ell}$ the level of branch length anisotropy, and $\theta_{\ell}$ the preferential orientation of the distribution. 
We observe that these distributions change their shape between cases A and B. 
While for case B, the shape of $\langle \ell \rangle(\theta)$ is evenly ellipsoidal, in case A we observe a larger proportion of branches pointing at $\theta \simeq 20^{\circ}$, while there is a decrease of branch lengths pointing at the orthogonal orientation $\theta \simeq 110^{\circ}$. 

\begin{figure}[htb]
  \centering
  \subfigure [ ] {
  \includegraphics[width=0.45\linewidth]{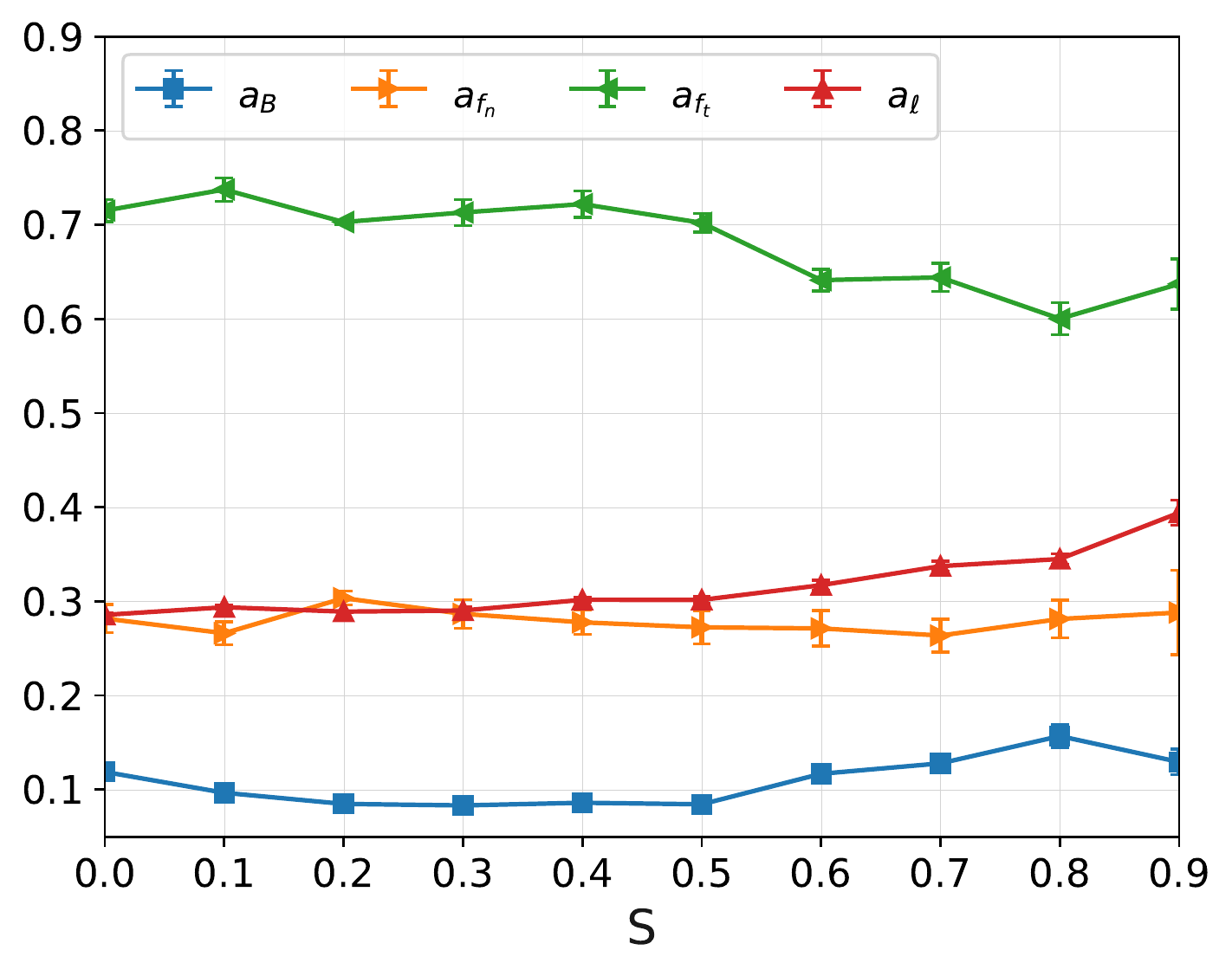}}
  \subfigure [ ] {
  \includegraphics[width=0.45\linewidth]{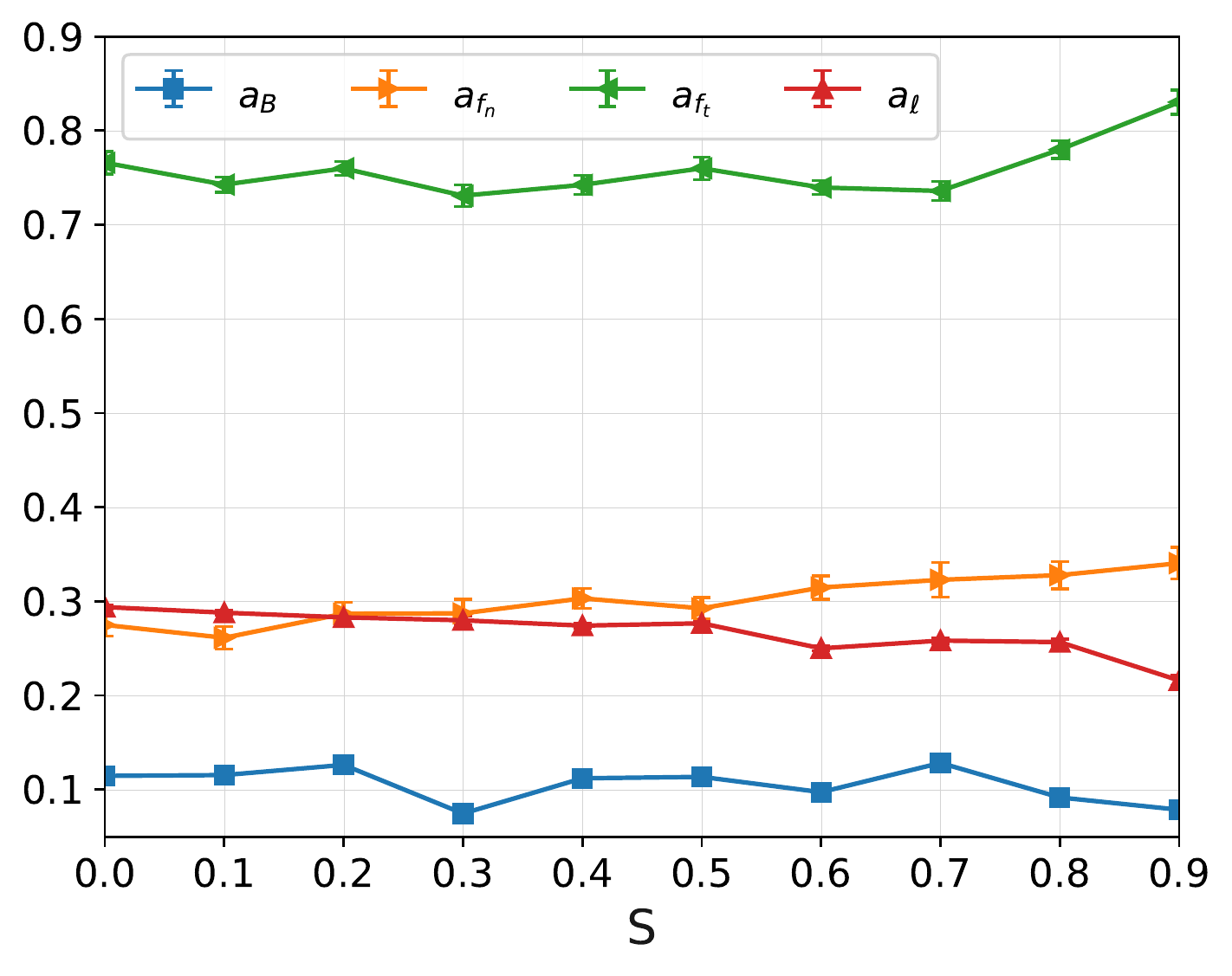}}
  \caption{Different anisotropies as a function of the grain size span $S$ for cases A (top) and B (bottom).}
  \label{fig:aniso}
\end{figure}

As mentioned above, simple fitting of the Fourier series to our data allows us to find all the different anisotropies and preferential orientation for the distributions of branch orientations, forces, and branch lengths. 
Figure \ref{fig:aniso} gathers the values of anisotropies for cases A and B as the particle size span increases. 
We observe that the anisotropies vary in a large range, spanning from $\simeq 0.1$ for the anisotropy of branch orientations to $\simeq 0.7$ for the ortho-radial force anisotropy. 

For case A, the branch orientation anisotropy $a_{B}$ and the branch length anisotropy $a_\ell$ seem to increase in similar proportions with $S$. 
This behavior is expected since an increase of grain size span also promotes contacts of different shape classes and more variability of branch lengths. 
On the other hand, the radial and ortho-radial force anisotropies do not seem to be affected by grain size span. 
For case B, the evolution of anisotropies is quite different. 
While the branch orientation anisotropy is roughly constant with $S$, the branch length anisotropy decreases with grain size span. 
The radial force anisotropy gradually increases with $S$ while the ortho-radial force anisotropy remains almost constant and only increases after $S>0.7$. 

Granular materials in which the particles are regular in size and shape keep particular relations between the main orientations of the different angular distributions, in the form $\theta_{f_n} \simeq \theta_{f_t} \simeq \theta_{B}$ and $\theta_{\ell'} \simeq \theta_{B} - \pi/2$. 
Using these relations, it is possible to deduce a well-known micromechanical expression for the shear strength as \cite{Rothenburg1989}
\begin{equation}\label{eq_notfull}
q/p \simeq \frac{1}{2} \{a_{B} + a_{f_n} + a_{f_t} + a_{\ell}\}.
\end{equation}

\begin{figure}[htb]
  \centering
  \subfigure [ ] {\includegraphics[width=0.45\linewidth]{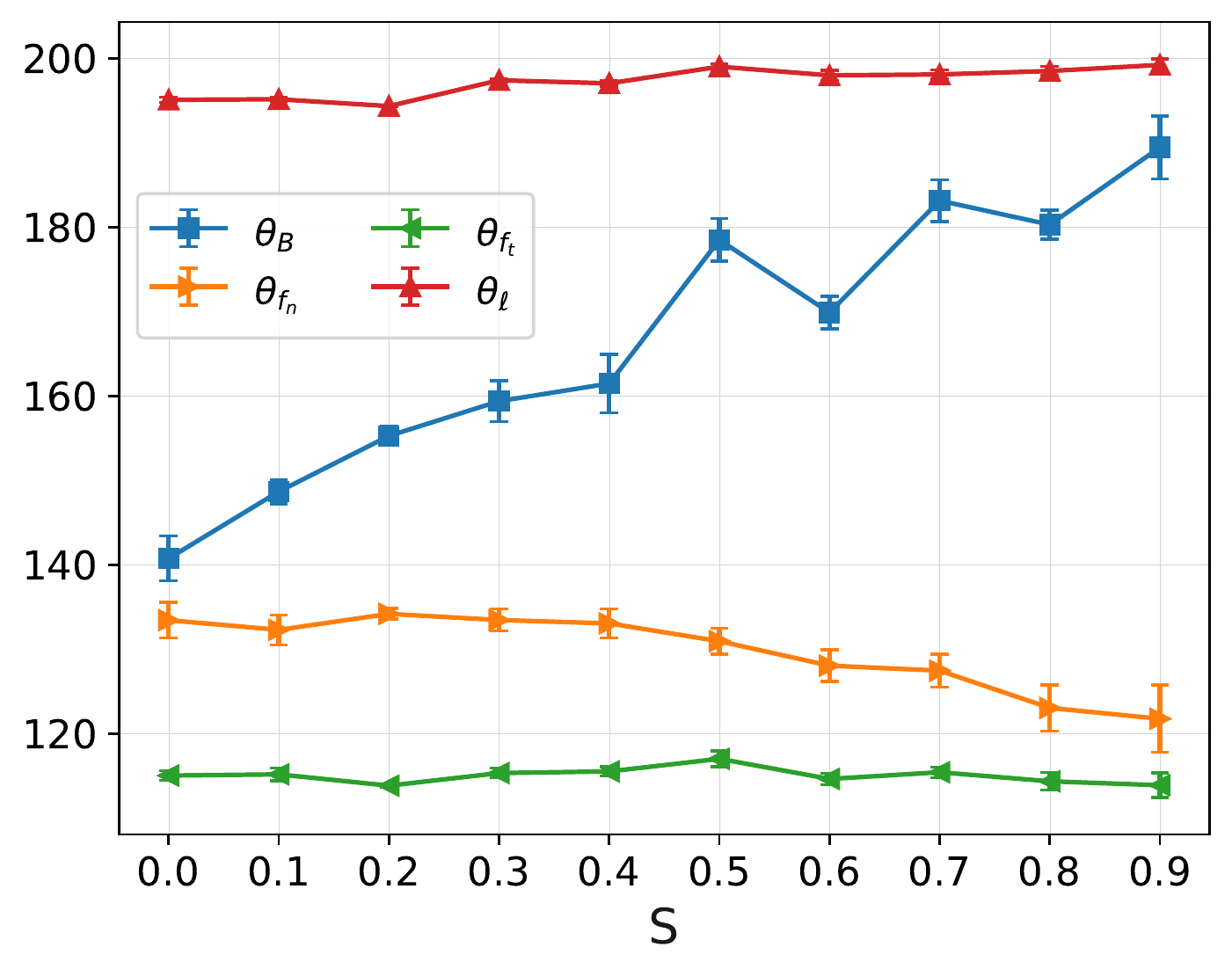}}
  \subfigure [ ] {\includegraphics[width=0.45\linewidth]{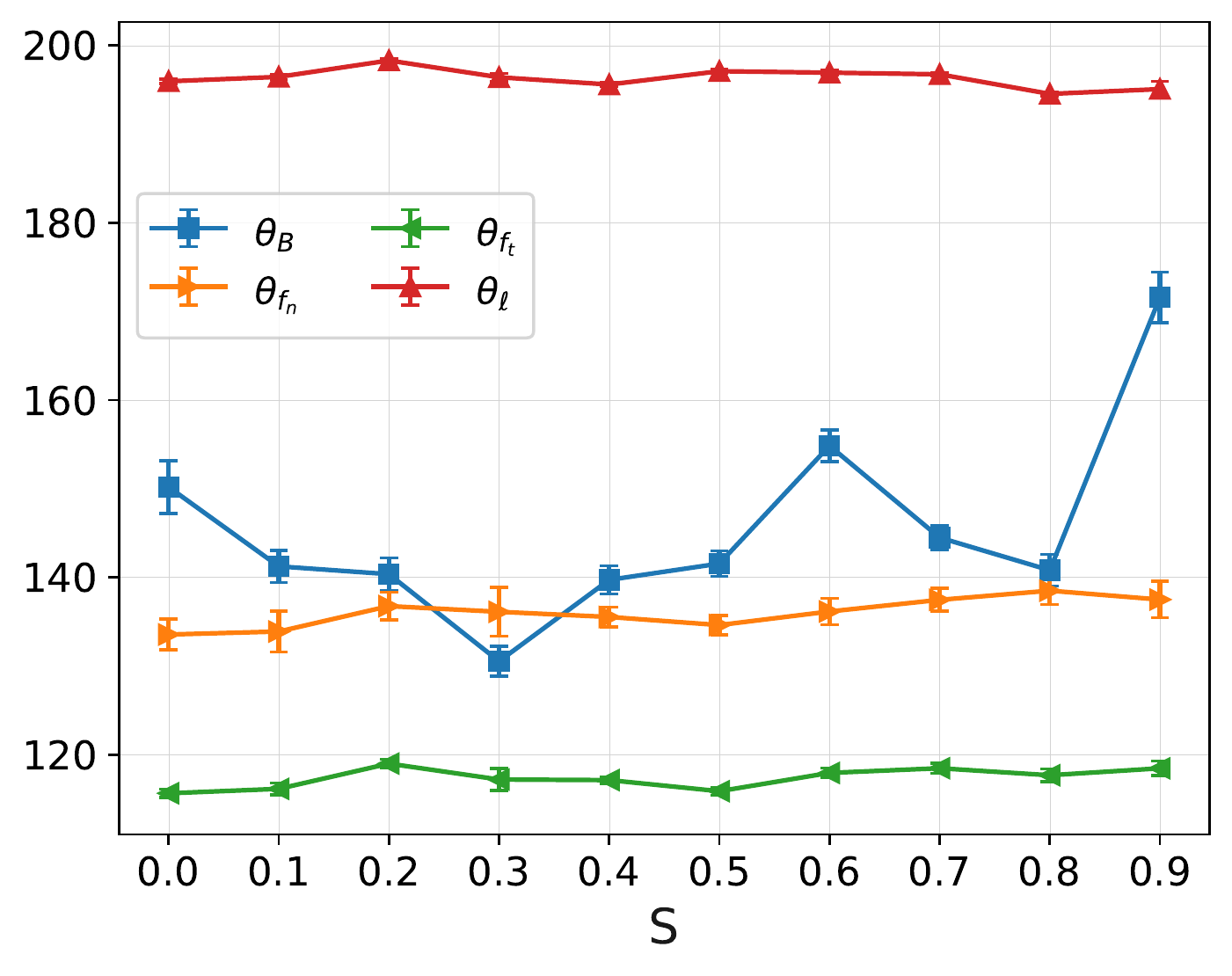}}
  \caption{Preferential orientations for the different angular distributions as a function of the grain size span $S$ for cases A (top) and B (bottom).}
  \label{fig:aniso_theta}
\end{figure}

However, the alignment of different angular distributions is broken when considering elongated particles and no assumptions can be made on the preferential orientations $\theta_B$, $\theta_{f_n}$, $\theta_{f_t}$, and $\theta_{\ell}$. 
To illustrate this fact, Fig. \ref{fig:aniso_theta} gathers all the preferential orientations as a function of $S$ for cases A and B. 
Effectively, no evident alignment occurs and the preferential orientations of branches and radial forces. 
In the meanwhile, the ortho-radial forces and branch main orientations seem independent of $S$. 

Under no assumptions for the preferential orientations of the angular distributions, we can analytically deduce a general expression to describe $q/p$. 
Let us write the stress tensor $\sigma_{ij}$ in integral form as \cite{Azema2010}

\begin{equation}\label{eq_full}
  \sigma_{ij} = n_c \int_{0}^{2\pi} P_B(\theta) \{\langle f'_n \rangle(\theta) n'_i +  \langle f'_t \rangle(\theta) t'_i \} \{\langle \ell \rangle(\theta) n'_j\} d \theta,
\end{equation}

where $n_c=Z/V$ is the volumetric contact density. 
This equation is very useful in this form since it involves the set of angular distributions of branches and forces we previously characterized.
Considering that the stresses $p$ and $q$ can also be written as 
\begin{equation}
p = \frac{\sigma_{11} + \sigma_{22}}{2},
\end{equation}

and 

\begin{equation}
q = \sqrt{\left\{\frac{1}{2} (\sigma_{11} - \sigma_{22})\right\}^2 + \sigma_{12}^2},
\end{equation}

we find the different components of $\sigma_{ij}$ integrating Eq. (\ref{eq_full}), neglecting higher-order terms and crossed products of different anisotropies.
We then deduce that the shear strength in the absence of any supposition upon the preferential orientations of the angular distributions as 
\begin{equation}\label{eq:aniso}
q/p \simeq \frac{1}{2} \sqrt{a_{B}^2 + a_{f_n}^2 + a_{f_t}^2 + a_{\ell}^2}.
\end{equation}

Despite the apparent similarity of this equation to Eq. (\ref{eq_notfull}), we should underline the fact that they are fundamentally different, deduced upon different suppositions, and should not be used arbitrarily. 
They do depend on the microstructure. 

\begin{figure}[htb]
\centering
\includegraphics[width=0.5\linewidth]{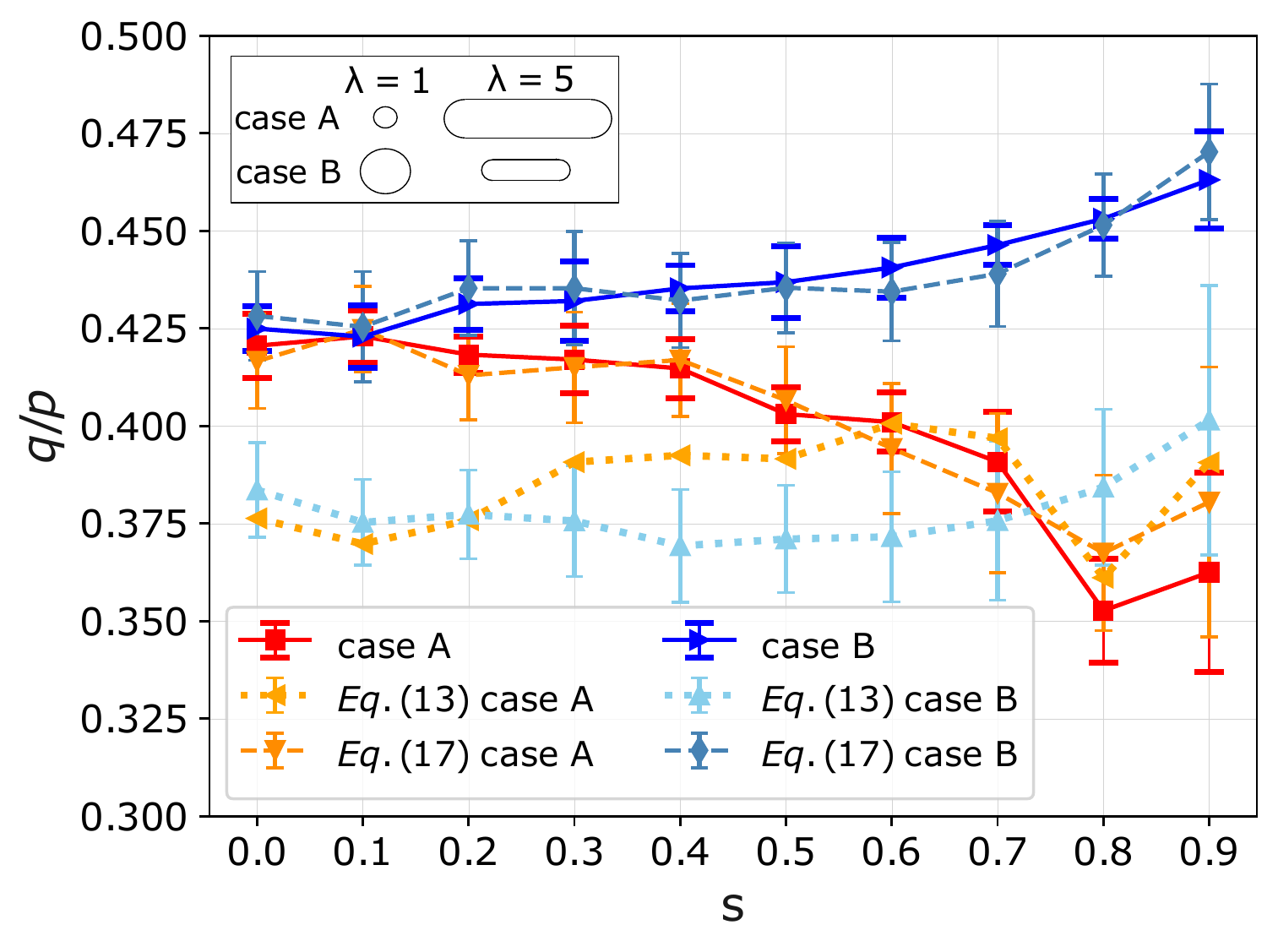}
\caption{Microstructural prediction of the shear strength $q/p$ by means of anisotropies using Eq. (\ref{eq_notfull}) and Eq. (\ref{eq:aniso}) along with the macroscopic values measured on Fig. \ref{fig:ss_strength}.}
\label{fig:comparison}
\end{figure}

In Fig. \ref{fig:comparison}, we plot the macroscopic shear strength observed in Fig. \ref{fig:ss_strength} and its prediction using both Eqs. (\ref{eq_notfull}) and (\ref{eq:aniso}) for cases A and B as a function of grain size span $S$. 
We see that Eq. (\ref{eq_notfull}) mismatches the measured shear strength. 
In turn, Eq. (\ref{eq:aniso}) fits very well the values of $q/p$. 

Then, we can understand that the drop of shear strength in case A is mainly supported by the drop of ortho-radial force anisotropy, despite increments of branch length and branch orientation anisotropies. 
This means that these samples are subjected to larger variations of local shear forces between grains. 
As shown by \cite{Azema2010,Cantor2018}, the friction mobilization scales with $a_{f_t}$. 
So, samples in case A are potentially mobilizing more friction due to kinematic restrictions that prevent the sample from deforming through the rotation of the bodies. 
In case B, despite a drop of branch length anisotropy, the increments of radial force anisotropy and the slight increment of ortho-radial force anisotropy are compensating mechanisms behind the almost independence of $q/p$ with $S$, which is in the same vein of many other numerical experiments investigating the shear behavior of regular size and shape grains. 
This opposite behavior between size-shape correlations A and B suggests that particle shape is a parameter that should be carefully analyzed before proposing a specific particle size scaling method.
Moreover, identifying preferential shapes for each particle size can be fundamental for each scaling approach. 

Since the role of each particle shape in a given psd can largely vary for a given granular material. 
We can decompose the total shear strength $q/p$ in order to assess the contributions of each shape class on the shear strength, as \cite{Cantor2020}
\begin{equation}
q/p = \frac{1}{p} \sum_{i=1}^{N_{sc}} q_i,
\end{equation}

where $q_i$ is the deviatoric component of each size class ($sc$). 
The corresponding value $q_i$ is found using the granular stress tensor (Eq. (\ref{eq_stresstensor})), but taking into account only the grains belonging to a particular $sc$. 

\begin{figure}[htb]
  \centering
  \subfigure [ ] {
  \includegraphics[width=0.45\linewidth]{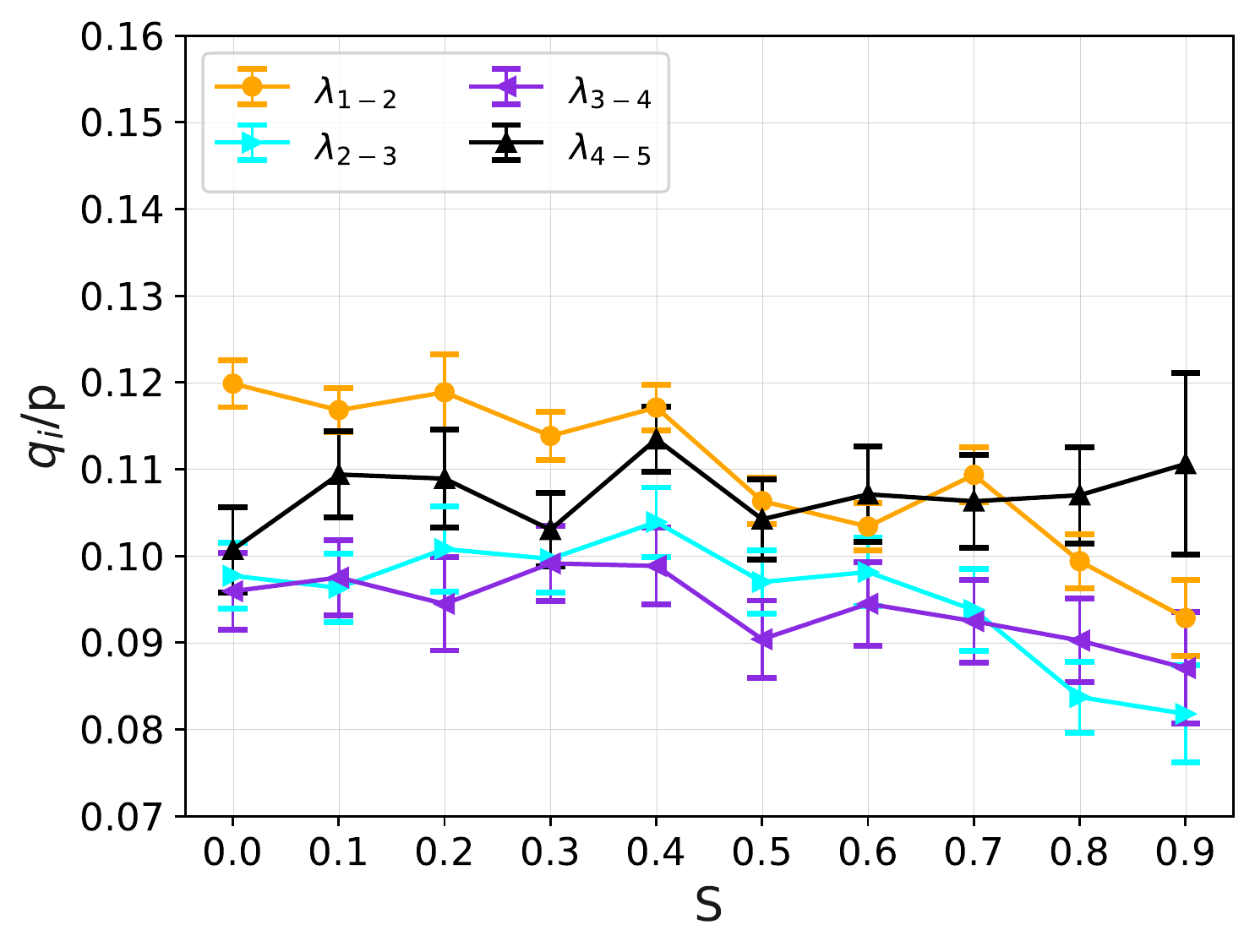}}
  \subfigure [ ] {
  \includegraphics[width=0.45\linewidth]{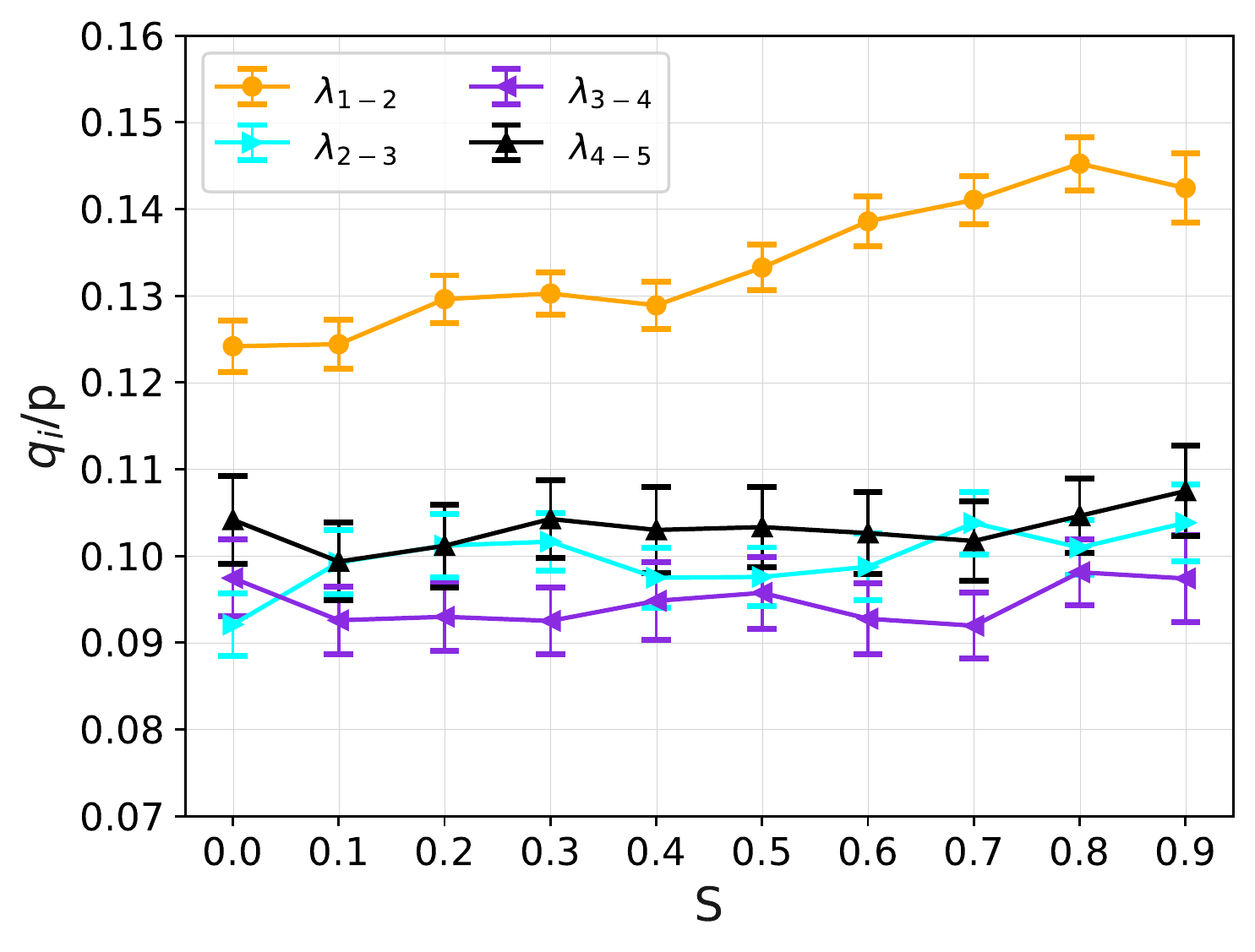}}
  \caption{Decomposition of the deviatoric component of stresses by particle shape $\lambda$ for cases A (a) and B (b) as a function of the grain size span $S$.}
  \label{fig:contrib_shapes}
\end{figure}

Figure \ref{fig:contrib_shapes} presents the deviatoric contributions of shape classes to $q/p$ for correlations A and B as a function of the increasing particle size span $S$. 
Note that we group the shape classes in intervals of $\lambda$ of 1. 
For instance, the subscripts in $\lambda_{1-2}$ mean that such a value gathers $q$ for the interval of particle elongations $\lambda \in [1,2[$. 

For case A, in which bigger particles are elongated, we expected that circular grains (i.e., $\lambda_{1-2}$) contribute relatively less to the shear strength as $S$ increases. 
This is indeed observed as the value $q_i$ for $\lambda_{1-2}$ decreases with $S$. 
The contributions of shape classes $\lambda_{2-3}$ and $\lambda_{3-4}$ also tend to decrease with $S$. 
On the other hand, the contributions of elongated big grains $\lambda_{4-5}$, remain almost independent of $S$ and, for large grains size spans $S$, they are indeed the particles contributing the most to the shear strength. 
These curves help to understand that the drop of $q/p$ in case A is due to a decrease of shear strength provided by the less elongated grains. 

For case B, while $q_i$ for $\lambda_{2-3}$, $\lambda_{3-4}$, and $\lambda_{4-5}$ are practically independent with $S$, the contribution of the more circular grains (i.e., $\lambda_{1-2}$) gradually increases with $S$ and remains as the main support to the shear strength among all the shape classes. 

\section{Conclusions}
In this paper, we studied granular assemblies in which correlations size-shape were considered in two scenarios: particle size distributions having A) large elongated and finer circular grains and B) large circular grains and finer elongated grains. 
Both correlations can be linked to distinctive geological formations and rock genesis. 
Our granular assemblies varied in grain size span from monosize to polydisperse systems in which the ratio $d_{max}/d_{min} = 19$, and the elongation of particles reached an aspect ratio of 5. 
Using a discrete-element approach, we tested these granular assemblies under periodic quasi-static shear conditions up to a cumulated shear deformation $\gamma = 400\%$. 

We characterized the steady shear strength and packing fraction as a function of the grain size span. 
We observed that, contrary to results in literature in which the shear strength turns out to be independent of the grain size span, including size-shape correlations deeply modify the mechanical response of the assemblies. 
In particular, systems in which larger grains are elongated and small grains are circular show a drop of shear strength as the grain size span increases. 
The opposite correlation, in which larger grains are circular and small grains are elongated, showed that shear strength barely increases with grain size span. 
These observations point out the importance of considering the shape of the grains when dealing with scaling methods in geotechnical engineering. 

To understand the different behavior in cases A and B, we undertook a microstructural and micromechanical analysis starting with particle orientation and connectivity. 
We showed that the average number of contacts per grain or coordination number $Z$ evolves similarly to the shear strength. 
While $Z$ decreases with grain size span for case A, its value slowly increases for case B. 
This result is counterintuitive given that case A has big elongated grains that are supposed to accommodate many neighboring smaller grains around them. 
Despite this, given the uniform particle size distribution by volume fractions, those elongated big grains turn out to be just a few in number and, although locally they have greater connectivity, this is not translated to the averaged value of $Z$ at the macroscale. 
As shown in the literature, the proportion of floating particles $c_0$ was found to increase in case A with particle size span. 
Nonetheless, case B showed a more complex behavior in which the value of $c_0$ decreases with grain size span and finds a minimum value for $S\simeq 0.6$. 
In this case, $c_0$ is able to increase only when the size dispersion is $S>0.6$. 
This seems to be related to the capacity of the bigger grains to create large enough voids for the smaller grains to rattle inside. 
Thus, wider particle size distributions are necessary when materials have a correlation type B in order to increase both their density and proportion of floating particles at the steady state. 

Then, employing a decomposition of the granular stress tensor in terms of microstructural anisotropies, we found out that the anisotropy of tangential forces is the main element supporting the variations of $q/p$ at the macroscopic scale. 
This analysis had to include an extended and detailed characterization of contacts, force intensities, and branch lengths since elongated particles break many of the standard relations and simplifications that can be done when making this microstructural analysis with rounded or mono-sized grains. 

Nonetheless, the approach of decomposing $q/p$ - into geometrical and force anisotropies - is still robust enough to link the micro and macro scales. 
Finally, we also decomposed the deviatoric component of the shear strength $q/p$ by contributions of different shape classes. 
This analysis showed that larger particles tend to contribute more to the shear strength as the grain size span increases. 
This is less evident, though, when the larger particles are elongated. 
Rounded particles seem to have an important role in shear strength independently of their granulometric class. 
This fact suggests that scaling methods based on truncated psd can add important errors to the estimation of shear strength and parallel scaling methods should certainly not avoid the size-shape correlations when it is such that the large grains are rounded and the small are elongated. 

Future studies in this vein should focus on additional elements of realistic correlations in the shape and size properties of geomaterials including angularity, the fracture strength of the grains, or extending this work to three-dimensional grains. 
The validation of these numerical observations by means of experimental tests is also needed but seldom seen in the literature. 
The challenges of experimental testing to reproduce broad grain size distribution while taking into account the shape is still an open issue that calls for large-scale testing or alternative physical approaches. 

\section*{Acknowledgements}
This research benefited from the financial support of the Natural Sciences and Engineering Research Council of Canada (NSERC) [Ref. RGPIN-2019-06118], the Fonds de recherche du Québec - Nature et technologies (FRQNT) through the « Programme de recherche en partenariat sur le développement durable du secteur minier-II » [Ref. 2020-MN-281267] and the industrial partners of the Research Institute on Mines and the Environment (RIME) UQAT-Polytechnique (irme.ca/en).
The authors acknowledge Dr. Sandra Linero and Pr. Emilien Azéma for fruitful and supportive discussions. Numerical simulations were made possible through support from Compute Canada under the Resources for Research Groups 2021 program (Project ID 3604).

\nocite{*}
\bibliography{references}
\clearpage


\end{document}